\documentstyle[twocolumn,aps,prc,epsfig]{revtex}

\input epsf

\begin{document}

\draft


\title{Neutron Stars in a \\
Class of Non-Linear Relativistic Models}

\author{A.R. Taurines $^{1}$,
C.A.Z. Vasconcellos $^{1}$, M. Malheiro $^{2}$,
 M. Chiapparini $^{3}$}

\address{$^{1}$ Instituto de F\'{\i}sica, Universidade Federal do Rio
Grande do Sul, Porto Alegre, 91501-970, Brasil\\ $^{2}$ Instituto
de F\'\i sica, Universidade Federal Fluminense, Niter\'oi,
24210-340, Brasil \\ $^{3}$ Departamento de F\'{\i}sica Te\'orica,
Universidade do Estado do Rio de Janeiro, Rio de Janeiro,
20559-900, Brasil}

\date{September 2000}
\maketitle

\begin{abstract}
We introduce in this work a  class of relativistic models for
nuclear matter and neutron stars which exhibits a
parameterization, through mathematical constants, of the
non-linear meson-baryon couplings. For appropriate choices of the
parameters,   it recovers current QHD models found in the
literature: Walecka and the Zimanyi and Moszkowski models (ZM and
ZM3). For other choices of the parameters, the new models give
very interesting and new physical results.We have obtained
numerical values for the maximum mass and radius of a neutron star
sequence, red-shift, hyperon populations, radial distribution of
particles, among other relevant static properties of these stellar
objects. The phenomenology of neutron stars in ZM models is
presented and compared to the phenomenology obtained in other
versions of the Walecka model. We have found that the ZM3 model is
too soft and predicts a very small maximum neutron star mass,
$\sim0.72 M_\odot$. A strong similarity between the results of
ZM-like models and those with exponential couplings is noted.
Sensibility of the results to the specific choice of the values
for the binding energy and saturation density is pointed out.
Finally, we discuss the very intense scalar condensates found in
the interior of neutron stars which may lead to negative effective
masses.
\end{abstract}


\narrowtext

\section{Introduction}
Since the dis\-co\-ve\-ry of the first pul\-sar in 1968
\cite{hewish} and its i\-den\-ti\-fi\-ca\-tion as a ro\-ta\-ting
neu\-tron star, the struc\-tu\-re, com\-po\-si\-tion, dy\-na\-mics
and e\-vo\-lu\-tion of these as\-tro\-phy\-si\-cal o\-bjects
be\-ca\-me im\-por\-tant the\-mes of theo\-re\-ti\-cal and
phe\-no\-me\-no\-lo\-gi\-cal re\-search. According to an early
suggestion \cite{baade}, neutron stars evolve from an initially
hot protoneutron star that forms in the collapse of a massive star
in the supernova phenomenon. At densities exceeding that of
nuclear matter, important static properties of a neutron star as
the mass-radius relation, the crust extent, the distribution of
the stellar moments of inertia and the central density, may be
determined by its equation of state (EOS) \cite{shapiro}.

During this period, there has been a continuous enhancement
concerning relativistic microscopic calculations of the EOS of
neutron stars, improving our understanding of the structure of
these stellar objects. In particular, any theory must at least
account for a neutron star as massive as the most massive observed
pulsar; the knowledge of pulsar masses provides a very important
constraint on the theory.

More recent calculations based on relativistic properties of
nuclear matter at high densities indicate that the equations of
state are considerable stiffer than those predicted by
non-relativistic approaches. As a result, the mass of a neutron
star is believed to be at least as large as $1.6-2.1M_{\odot}$.

From the theoretical point of view, quantum chromodynamics (QCD)
represents the most profound description of the strong interaction
and would be the ideal tool for neutron star applications.
However, the highly non-linear behavior of QCD at the hadronic
energy scales inhibits any practical calculations leading most
theorists to search for phenomenological descriptions of the
structure of nuclear matter. One of these alternative approachs is
quantum hadrodynamics (QHD) \cite{walecka 86}, a relativistic
quantum field theory based on a local Lagrangian density which
uses ba\-ry\-on and me\-son fields as the re\-le\-vant degrees of
freedom.
This model provides consistent theoretical framework for
describing such a relativistic interacting many-body system and,
based on it, Glendenning presented \cite{glendenning 85} a very
comprehensive treatment of the matter in neutron stars using an
extended version which included leptons and the fundamental
baryonic octet.

Alternative versions  of the Walecka model, namely the
Boguta-Bodmer (BB)  \cite{boguta} and the Zimanyi-Moszkowski (ZM)
models \cite{zimanyi,delfino}, were developed to improve the
description of the nucleon effective mass, $M^\star$, (too low)
and compression modulus of nuclear matter, $K$, (too high) as
attained with the original approach. Boguta and Bodmer introduced
cubic and quartic scalar self-interactions in the Lagrangian while
the ZM models have heightened the delineation of these quantities
by
replacing 
the Yukawa scalar coupling term 
by a {\em derivative coupling} contribution. This derivative
coupling may be interpreted alternatively as a phenomenological
coupling between the scalar neutral meson and the nucleon fields
through the introduction of a baryon density dependence in the
scalar, vector and isovector coupling constants of the theory
\cite{Chiapparini}.

In this work we analyze the structure of neutron stars by
introducing a QHD Lagrangian with a parameterized meson coupling
contribution \cite{taurines}. This phenomenological approach
contains high order self-coupling contributions of the meson
fields and permits in particular to restore the results obtained
with Walecka, ZM  and ZM3 models by making suitable choices of the
values of the mathematical parameters of the theory. The control
on the analytical form of the couplings allows us to investigate
other values of these parameters which give new  physical results.
By extending the formalism to include hyperons and leptons, we
investigate several static bulk properties of neutron stars using
the Walecka, ZM and ZM3 models, the two later being applied to
this problem for the first time. As we have a class of models we
are able to relate nuclear matter and neutron star properties. In
particular, we have found that some of the studied models describe
very strong scalar fields in the interior of the neutron stars,
leading to a negative nucleon effective mass.

\section{General Characteristics of
Neutron Stars  \label{3seccar}}

\subsection{Electrical Charge Neutrality}
According to Glendenning in \cite{glendenning 85}, {\em la raison
d'\^etre} of neutron stars is to be neutral. To prove this
assertive, one should consider that neutron stars are held
together by the gravitational attraction and take into account the
balance between the repulsive Coulomb force acting on a charged
particle of the same sign as the net charge of the star
($Z_{net}$) and the gravitational force. Assuming the particle to
be located at the surface of the star, it will be expelled out
unless the gravitational force overcomes the Coulomb force. For
the proton, the corresponding limit on the net positive charge of
the star is $Z_{net}/A\sim 10^{-36}$ \cite{glendenning 97}. For
the electron, this limit would be reduced by the factor $m_e
/m_p$. Hence, the net charge per charged particle is practically
zero. This result leads to the conclusion that a neutron star is
electrically neutral.

\subsection{Chemical Equilibrium}
In the evolution of the protoneutron star many different reactions
can occur. Electric charge and baryon number are conserved on a
long time-scale in comparison to the lifetime of the star. In the
core of the protoneutron star, the Fermi energy of the nucleons
exceed the hyperon masses and these particles can be produced in
strong interaction processes with conservation of strangeness in
reactions such that
\begin{eqnarray}
n+n \rightarrow n+\Lambda+K  . \label{eqlamb}
\end{eqnarray}
However, strangeness is not conserved in the time-scale of the
star since there occurs diffusion of neutrinos and photons to the
surface of the star  and processes like \begin{eqnarray}
 K^0  & \rightarrow &  2 \gamma  \, \,  ;  \, \, \, \,
K^-  \rightarrow  \mu^- + \bar{\nu}_{\mu} ,...    \label{eqkaon}
\end{eqnarray}
cannot be reversed anymore and a net strangeness appears.

In this evolution process, the star reaches the chemical
equilibrium, a degenerate state where, from the point of view of
its hadronic and leptonic composition, further reactions are not
possible. As an example, in an ideal degenerate system of protons,
neutrons and electrons at chemical equilibrium, particle levels
are filled in such a way that neutron beta decay or proton inverse
beta decay are not energetically favored.

In general, if one takes into account the fundamental baryon octet
and lepton degrees of freedom (electrons and muons), the following
chemical equilibrium equation then holds (see Appendix
\ref{appendixa})
\begin{eqnarray} \mu_i = q_{b,i}\mu_n - q_{e,i}\mu_e \, ,
\end{eqnarray}
where $i = p, n, \Lambda, \Sigma^0, \Sigma^-, \Sigma^+, \Xi^-,
\Xi^0, e, \mu$; $q_{b,i}$ represents the baryon number and
$q_{e,i}$ the electric charge of species $i$. In this way, the
conditions for $\beta$ equilibrium can be summarized as
\begin{eqnarray} \mu_{\Sigma^0} & = &
\mu_{\Xi^0}= \mu_{\Lambda}  =   \mu_n \, ;  \nonumber \\
 \mu_{\Sigma^-} &  = &
\mu_{\Xi^-}=\mu_n+\mu_e \, ;   \\ \mu_{\Sigma^+} & = &  \mu_p =
\mu_n-\mu_e
 \, . \nonumber
\end{eqnarray}

\section{Boguta-Bodmer  Model} \label{3secpop}
\subsection{Theory}
In this section we study  baryon and lepton populations in neutron
stars by using the Boguta-Bodmer model (BB) with hyperon degrees
of freedom. In spite of this study having been already done
\cite{glendenning 85}, we reproduce its main results as a guide
for the development in the next section of our new class of
non-linear relativistic models.

The BB model describes the complex fermionic composition of
neutron stars as a generalization of the $\sigma$, $\omega$ and
$\varrho$ theory
\begin{eqnarray}
 {\cal{L}}   & \! \! = \! \!  & \sum\limits_{B}
\bar{\psi}_B [i \gamma_\mu \partial^\mu
\! - \!  (M_B \! - \!
g_{\sigma B} \sigma) \! - \! g_{\omega B} \gamma_\mu \omega^\mu] \psi_B
\nonumber \\ & \! \!  - \! \! &
\sum\limits_{B} \bar{\psi}_B [    \frac12 g_{\varrho B}
\gamma_\mu \mbox{\boldmath$\tau$}\cdot \mbox{\boldmath$\varrho$}^\mu]
\psi_B
\! \! +  \! \!    \frac12(\partial_\mu
\sigma \partial^\mu \sigma - {m_\sigma}^2 \sigma^2) \nonumber \\
 & \! \! - \! \! &
\frac14 \omega_{\mu \nu} \omega^{\mu \nu}  +   \frac12 {m_\omega}^2
\omega_\mu \omega^\mu
\! \!  - \! \!      \frac14 \mbox{\boldmath$\varrho$}_{\mu \nu}\cdot
\mbox{\boldmath$\varrho$}^{\mu \nu}
  \! \! + \! \!     \frac12m_\varrho^2 \mbox{\boldmath$\varrho$}_\mu \cdot
\mbox{\boldmath$\varrho$}^\mu  \nonumber \\
& \! \!  - \! \!  & \frac13bM(g_\sigma\sigma)^3
 -\frac14c(g_\sigma\sigma)^4
\! \!  + \! \! \sum\limits_\ell
\bar{\psi}_\ell(i\gamma_\mu\partial^\mu- m_\ell)\psi_\ell \, .
\nonumber \\  \label{nltot}
\end{eqnarray}
This Lagrangian density describes a system of eight baryons ($B =
$ $p$, $n$, $\Lambda$, $\Sigma^-$, $\Sigma^0$, $\Sigma^+$,
$\Xi^-$, $\Xi^0$) coupled to three mesons ($\sigma$, $\omega$,
$\varrho$) and two free lepton species ($\ell = e^-, \mu^-$).
The scalar and vector coupling constants in the theory,
$g_{(\sigma,\omega)}$, and the coefficients $b$ and $c$ are
determined to reproduce, at saturation density $\rho_0 = 0.15
fm^{-3}$, the binding energy of nuclear matter, $B = -16 MeV$, the
compression modulus of nuclear matter, $K=250 MeV$, and the
nucleon effective mass, $M^\star/M=0.75$. In fact, the values for
these two last quantities are not well established and we have
just taken the most used values in the literature of the BB model.
Additionally, to describe the symmetry energy coefficient, $a_4 =
32.5 MeV$, we determine the isovector coupling constant
$g_{\varrho}$. We have found 
\begin{eqnarray} \left(\frac{g_{\sigma}}{m_{\sigma}}\right)^2 & =
& 9.86 \, fm^2 \, \, ; \, \,
\left(\frac{g_{\omega}}{m_{\omega}}\right)^2 = 5.85 \, fm^2 \, \,
; \nonumber \\ \left(\frac{g_{\varrho}}{m_{\varrho}}\right)^2 & =
& 4.80 \, fm^2 \, ; \,\, b= 0.00103 \, ; \,\, c=0.0100 .
\end{eqnarray}
{In the comparison with the results obtained by \cite{glendenning
85}, one should recall that this author has fitted the coupling
constants of the theory with $B=-15.95 MeV$, $\rho_0 = 0.145
fm^{-3}$,$\,K=285 MeV$, $M^\star/M=0.77$ and $a_4=36.8 MeV$.}

Using the Euler-Lagrange equations, the 
Dirac equation for uniform matter, in momentum representation, is
\begin{eqnarray} \left[\gamma_\mu (k^\mu
\!-\!g_{\omega B}\omega^\mu\! -\! \frac12 g_{\varrho B}
\mbox{\boldmath$\tau$}\cdot\mbox{\boldmath$\varrho$}^\mu)\! -\!
M^*_B(\sigma) \right] \psi_B(k)=0 \, ,  \label{nltotd}
\end{eqnarray}
where $M_B^*(\sigma) \equiv M_B-g_{\sigma B} \sigma$ is the
effective mass of the baryonic  species $B$. Furthermore, by
applying the mean field approximation, the $\omega_0$,
$\varrho_{03}$ and $\sigma$ meson field equations for uniform
static matter are
\begin{eqnarray} \omega_0 & = & \sum\limits_B\frac{g_{\omega
B}}{m^2_\omega} \rho_B \, , \nonumber 
\\
\varrho_{03} & = & \sum\limits_B \frac{g_{\varrho
B}}{m^2_\varrho}I_{3B} \rho_B \,\nonumber ,
\\ m^2_\sigma \sigma & = & -b M g_\sigma (g_\sigma\sigma)^2 -c
g_\sigma (g_\sigma \sigma)^3  \nonumber \\ & + &   \sum\limits_B
\frac{2J_B+1}{2 \pi^2} g_{\sigma B} \int_0^{k_{F,B}}
\frac{M_B^*(\sigma)}{\sqrt{k^2+ M_B^{* \, 2}}}k^2 dk \, .
\label{sigma}
\end{eqnarray}
In these equations, the baryon source terms
have been replaced by their ground state
values.

Defining the ratio between meson-hyperon and meson-nucleon coupling
constants as
\begin{equation}
\chi_{(\sigma,\omega,\varrho),B} \equiv
\frac{g_{(\sigma,\omega,\varrho),B}}
{g_{(\sigma,\omega,\varrho)}},
\end{equation}
we have, from equations (\ref{sigma}),
\begin{eqnarray}
g_\omega \omega_0 & = & \left(\frac{g_\omega}{m_\omega}\right)^2
\sum\limits_B\chi_{\omega,B} \rho_B \, , 
\\
g_\varrho\varrho_{03}  & = &
\left(\frac{g_{\varrho}}{m_\varrho}\right)^2 \sum\limits_B
\chi_{\varrho,B} I_{3B} \rho_B \, , \\ 
g_\sigma\sigma & = & \left(\frac{g_\sigma}{m_\sigma}\right)^2 [-b
M (g_\sigma\sigma)^2 -c (g_\sigma \sigma)^3 \nonumber \\ & + &
\sum\limits_B \frac{ \chi_{\sigma B}}{2 \pi^2} \int_0^{k_{F,B}}
\frac{M_B^*(\sigma)}{\sqrt{k^2+ M_B^{* \, 2}}}k^2 dk] \,
\label{sigmamf}.
\end{eqnarray}

The corresponding equations for baryon number and electric charge
conservation are: \begin{eqnarray} \rho = \sum\limits_B
\frac{k_{F,B}^3}{3 \pi^2} \, ,
\end{eqnarray}
and \begin{eqnarray} \sum\limits_B q_{e,B}\frac{k_{F,B}^3}{3
\pi^2} - \sum\limits_{\ell} \frac{k_{F,\ell}^3}{3 \pi^2}=0 \, .
\end{eqnarray}
The baryon chemical potentials, $\mu_B(k)$, correspond to
eigenvalues of the Dirac equation (\ref{nltotd}): \begin{eqnarray}
\mu_B(k) = g_{\omega B} \omega_0 + g_{\varrho B}
\varrho_{03}I_{3B} +\sqrt{k_{F,B}^2 + M_B^*(\sigma)^2} \, .
\label{nlpot}
\end{eqnarray}
In this expression, $I_{3B}$ is the isospin projection
of baryon charge states $B$ and $k_{F,B}$ is the
Fermi momentum of species $B$.

The EOS is obtained from the ground state expectation value of the
time and space components of the diagonal energy-momentum tensor.
The energy density and pressure of the system are given, in the BB
model, by
\begin{eqnarray} \varepsilon & = & \frac13
bM(g_\sigma\sigma)^3+\frac14c(g_\sigma\sigma)^4 +\frac12
m_\sigma^2 \sigma^2 + \frac12m_\omega^2 \omega_0^2 \nonumber \\ &
+ & \frac12 m_\rho^2 \varrho_{03}^2 + \sum\limits_B
\frac{1}{\pi^2} \int_0^{k_{F,B}}\sqrt{k^2 + M^{* \, 2}_B}k^2dk
\nonumber \\ & + &  \sum\limits_{\ell} \frac{1}{\pi^2}
\int_0^{k_{F,\ell}} \sqrt{k^2 + m_\ell^2}k^2dk \, ;
\label{3nldener}    \\ p  & = &    -\frac13
bM(g_\sigma\sigma)^3-\frac14c(g_\sigma\sigma)^4 -\frac12
m_\sigma^2 \sigma^2 + \frac12 m_\omega^2 \omega_0^2 \nonumber \\ &
+ &   \frac12 m_\rho^2 \varrho_{03}^2
  +   \frac13 \sum\limits_B\frac{1}{\pi^2} \int_0^{k_{F,B}}
\frac{ k^4dk}{\sqrt{k^2 + M^{* \, 2}_B}}  \nonumber \\ & + &
\frac13 \sum\limits_\ell\frac{1}{\pi^2} \int_0^{k_{F,\ell}} \frac{
k^4dk}{\sqrt{k^2 + m_\ell^2}} \, . \label{3nlpressao}
\end{eqnarray}
We present in the following 
the results obtained with this model.

\subsection{ Results}

In the numerical calculations with the BB model we have considered
matter with and without hyperons, in order to understand how these
strange species affect the neutron star properties.

\subsubsection{Matter with nucleons, hyperons and leptons}
\begin{figure}
\caption{Fermionic populations and field strengths in the
BB model for matter with
hyperons (panels {\em a,b} correspond to $\chi=\sqrt{2/3}$ and
panels {\em c,d} to $\chi=1$) and without them (panels {\em e,f}).
\label{3Ipop1}}
\begin{tabular} {c c}
{a)} & {b)} \\  {\epsfysize=6cm
\epsfbox{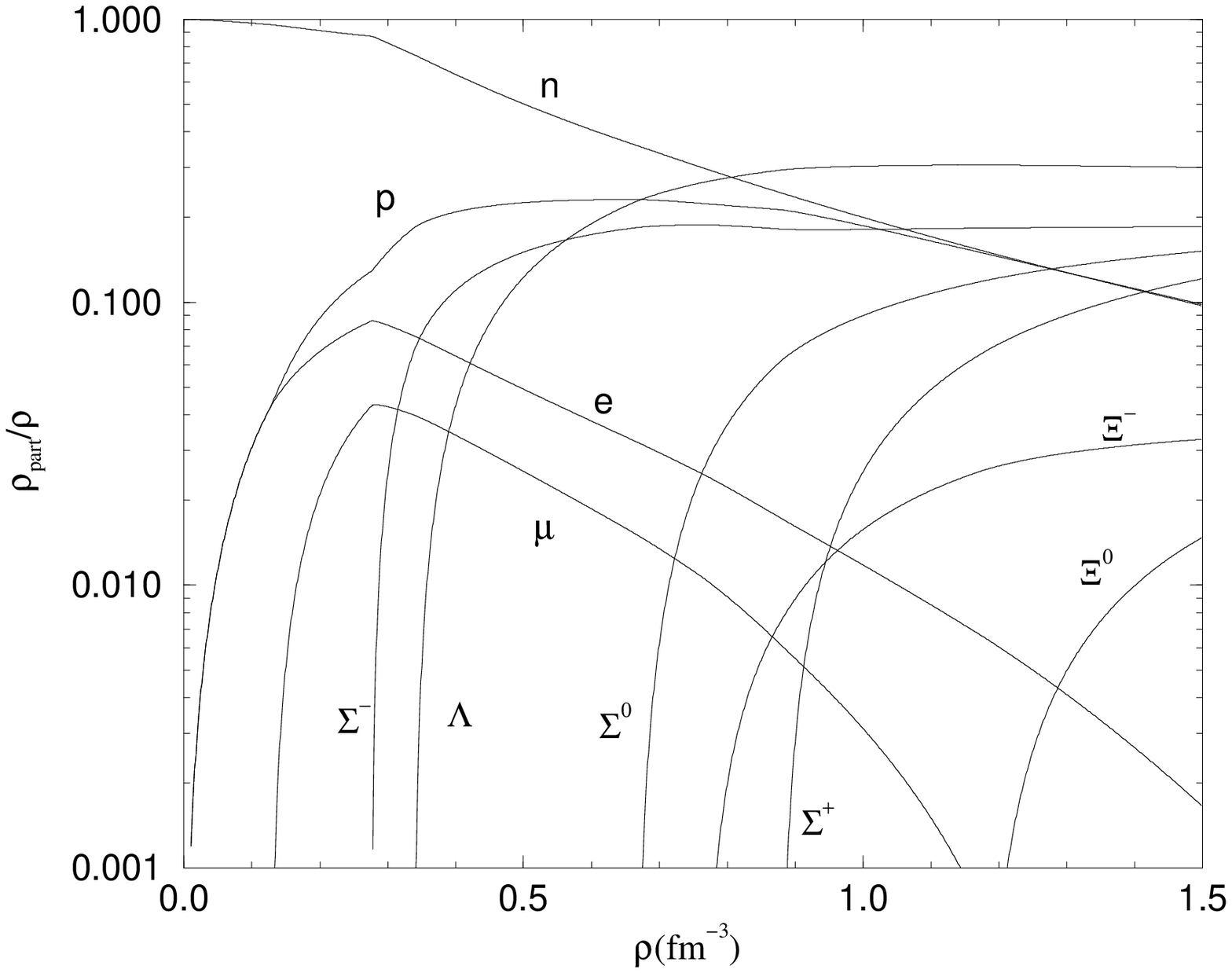}}&{\epsfysize=6cm \epsfbox{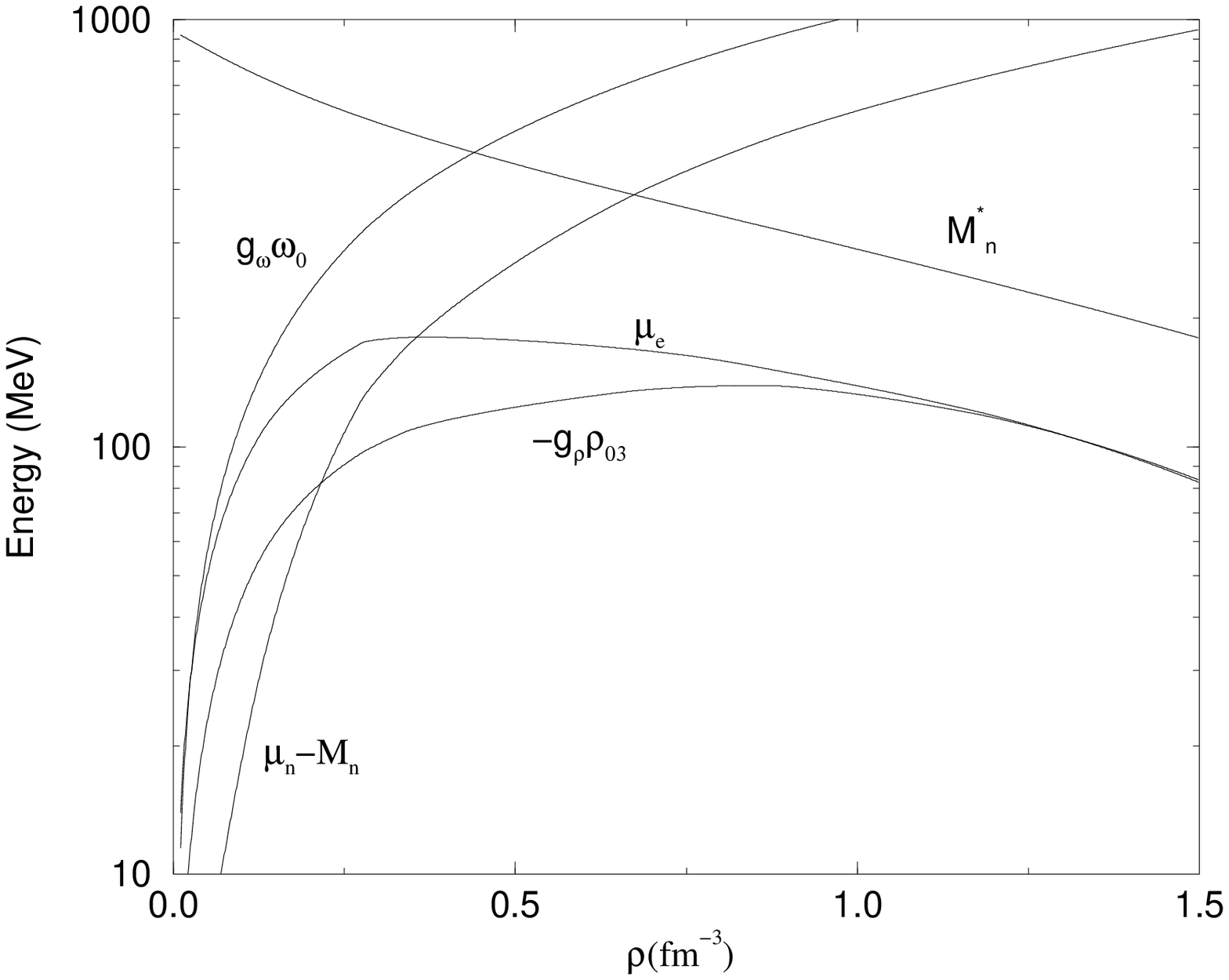}}\\
{c)} & {d)} \\  {\epsfysize=6cm
\epsfbox{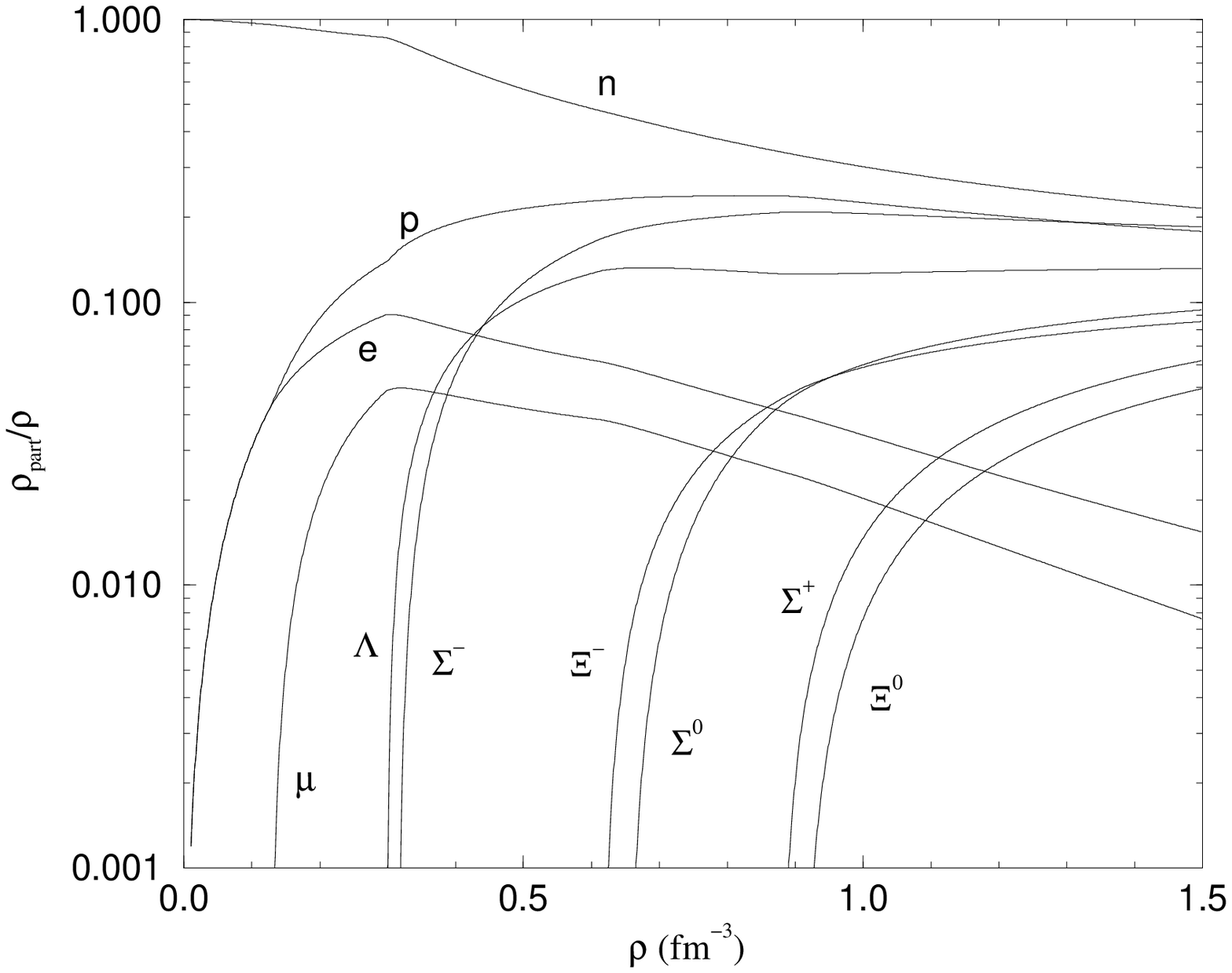}}&{\epsfysize=6cm \epsfbox{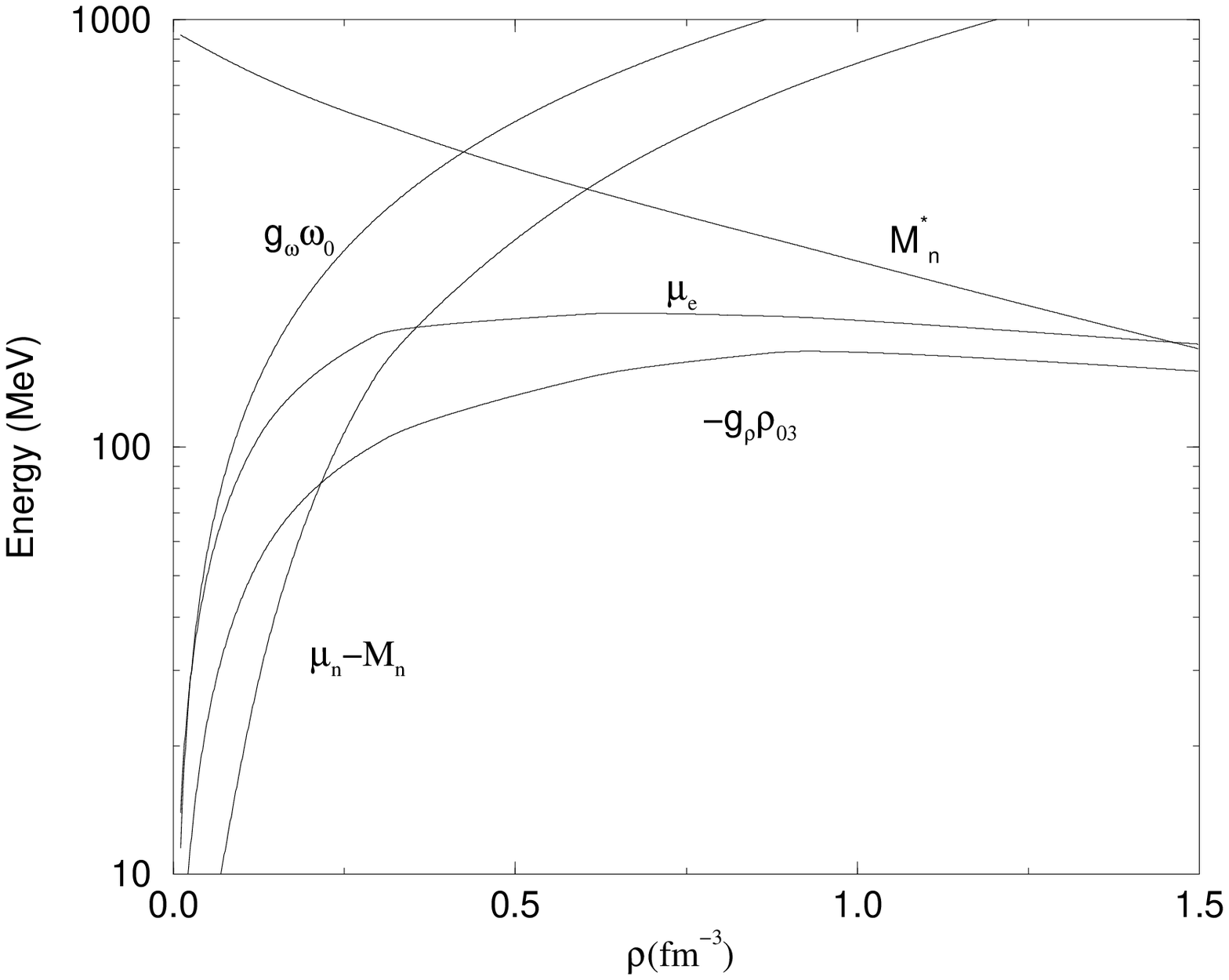}}\\
{e)} & {f)} \\  {\epsfysize=6cm
\epsfbox{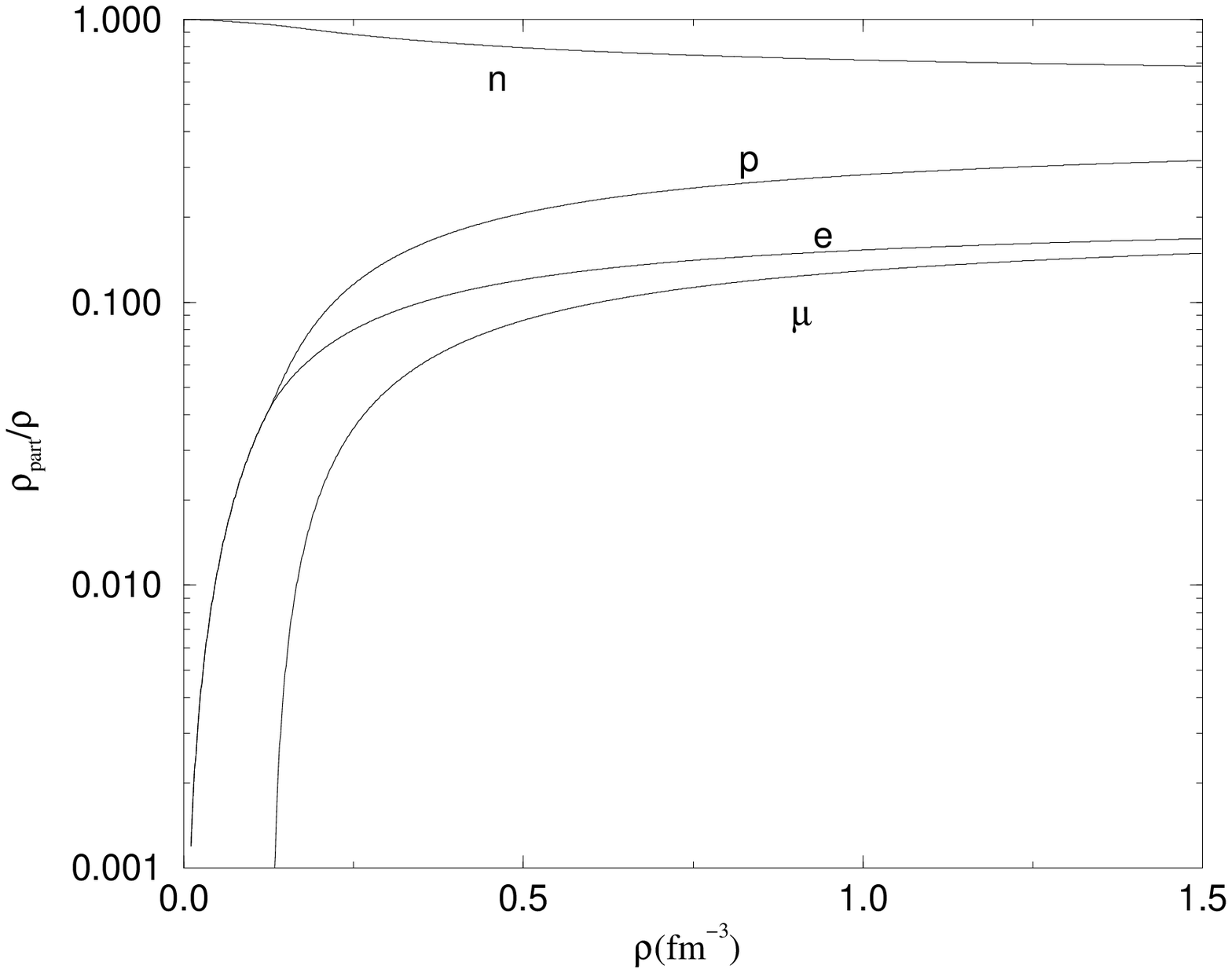}}&{\epsfysize=6cm \epsfbox{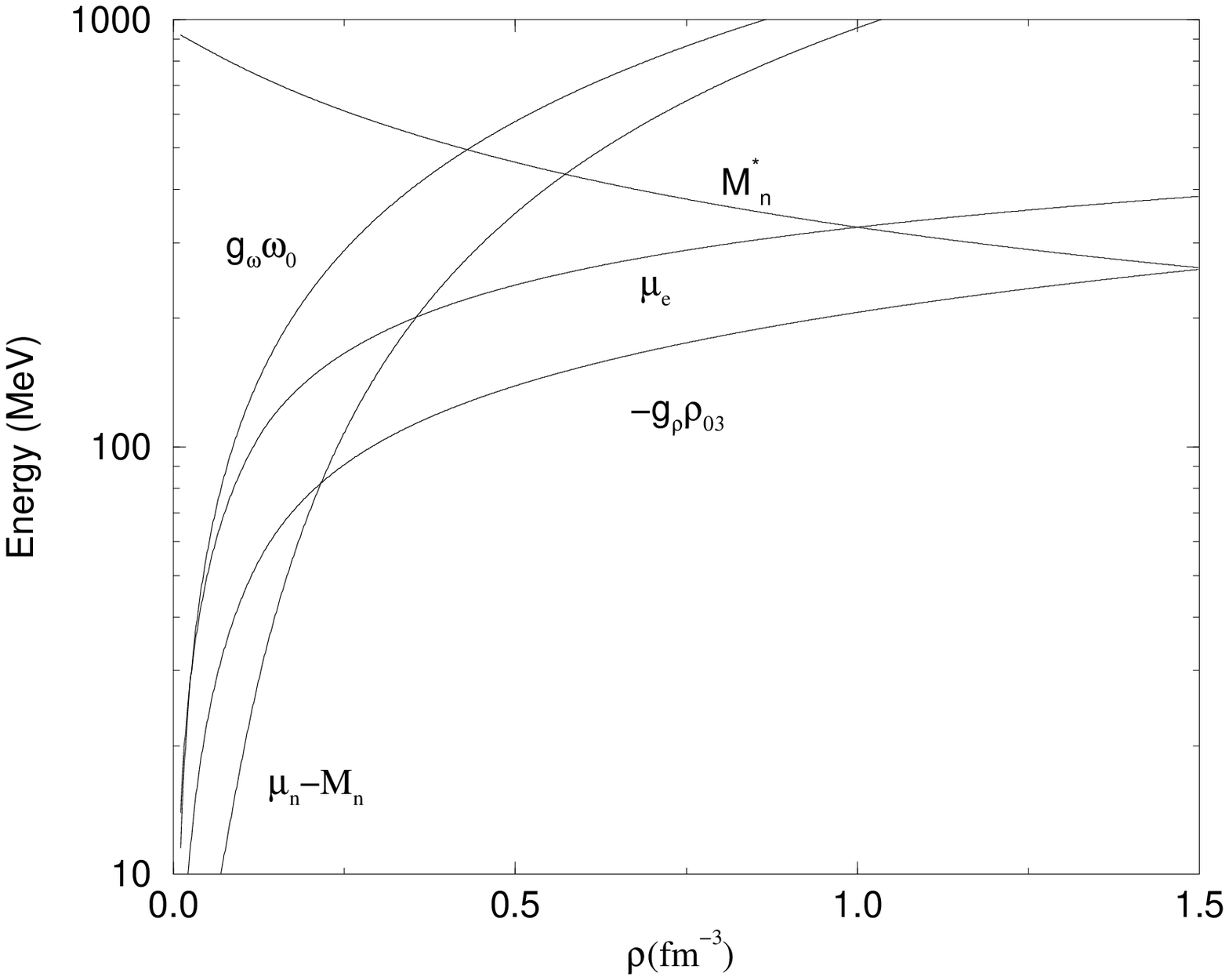}}\\
\end{tabular}
\end{figure}

Figure \ref{3Ipop1}, panels {\em a-d}, shows baryon and lepton
populations and field strengths as a
function of the total baryon density 
for two different ratios: $\chi=\sqrt{2/3}$ \footnote{This choice
is based on the quark counting of the baryons
\protect{\cite{mos-coup}}. } and $\chi=1$ (universal coupling).
From expression (\ref{nlpot}) for the baryon chemical potential we
can see that the charge term in the eigenvalue determines whether
a species is charge-favored or unfavored; and the isospin term
determines whether a species is isospin-favored or not. Baryons
with the same sign of the electric charge as the proton are
unfavored; baryons with the same sign of its isospin projection
are favored (notice that $g_{\varrho}\varrho_{03}<0$).

At high densities ($\rho\sim0.8 \, fm^{-3}$ ) the $\Lambda$
hyperon becomes the most populous species for the case
$\chi=\sqrt{2/3}$ (panel {\em a}). At panel {\em b} we see that
the electron chemical potential reaches a maximum value at $\sim
200\, MeV$ and begins to decrease due to the reduction of the
electron population. At $\rho\sim1.5\, fm^{-3}$ the nucleon
effective mass is still at $200 \, MeV$.

In the panel {\em c} of the same figure, we see that, to the
increasing of the meson-hyperon coupling constant (from
$\sqrt{2/3}$ to  $1$) it corresponds an early emergency of the
particles. In the comparison with the results of panel {\em a}, we
can see that the leptons have a greater population in this case
and that the neutron population remains always as the most
important in the system. The electron chemical potential and
$-g_\varrho{\varrho_{03}}$ saturates around $200\, MeV$ at panel
{\em d} and the nucleon effective mass behaves similarly to the
case $\chi=\sqrt{2/3}$.

\subsubsection{Matter with nucleons and leptons}
In the sequence of the analysis of the results of figure
\ref{3Ipop1}, panel {\em e} shows baryon and lepton populations
and the corresponding chemical potentials when we exclude hyperon
degrees of freedom. In this case, since charge neutrality is kept
only by the $p^+,e^-$ and $\mu^-$ particles, the lepton population
increases in the domain of densities shown in  the figure. Also,
we notice the increasing of the nucleon effective mass, which
means a less intense scalar field when compared to the previous
cases. This indicates that the {\em introduction of hyperons
enhance the scalar meson condensation}.

\subsubsection{Neutron star properties}
We are able now to find numerical results for the EOS using
equations (\ref{3nldener}) and (\ref{3nlpressao}). However, this
EOS corresponds to neutron star matter densities
($10^{13}-10^{15}g/cm^3$) and should be supplemented by EOS's from
other models for sub-nuclear densities: we adopt the approach
developed in \cite{harrison} in the density interval $2\times
10^3-1 \times 10^{11}g/cm^3$ and  in the range
$1\times10^{11}-2\times10^{13}g/cm^3$ we use the EOS presented in
\cite{vautherin 73}. Combining these EOS's with the
Tolman-Oppenheimer-Volkoff (TOV) equations
\cite{Tolman,Oppenheimer}, we may determine the mass of a neutron
star as a function of its central density.  To gain the
mass-radius relationship, the radius $R$ of a neutron star is
obtained with the condition that the pressure is null at the
surface of the star, $p(R) = 0$. We have found values for the mass
and radius of different neutron star sequences as a function of
the central density $\varepsilon_c$.

\begin{figure}
\caption{ {\em Panel a}: distribution of the energy density throughout a star
composed of nucleons and leptons. {\em Panel b}: Neutron star mass as a function
of the central density for matter with (curve I corresponds to $\chi=\sqrt{2/3}$ and curve II
to $\chi=1$) and without (curve III) hyperons. {\em Panel c}: mass-radius relation (same
labels as in panel b). 
 \label{starnl}}
\begin{tabular} {c}
{\footnotesize{a)}} \\  \centerline{\epsfysize=6cm
\epsfbox{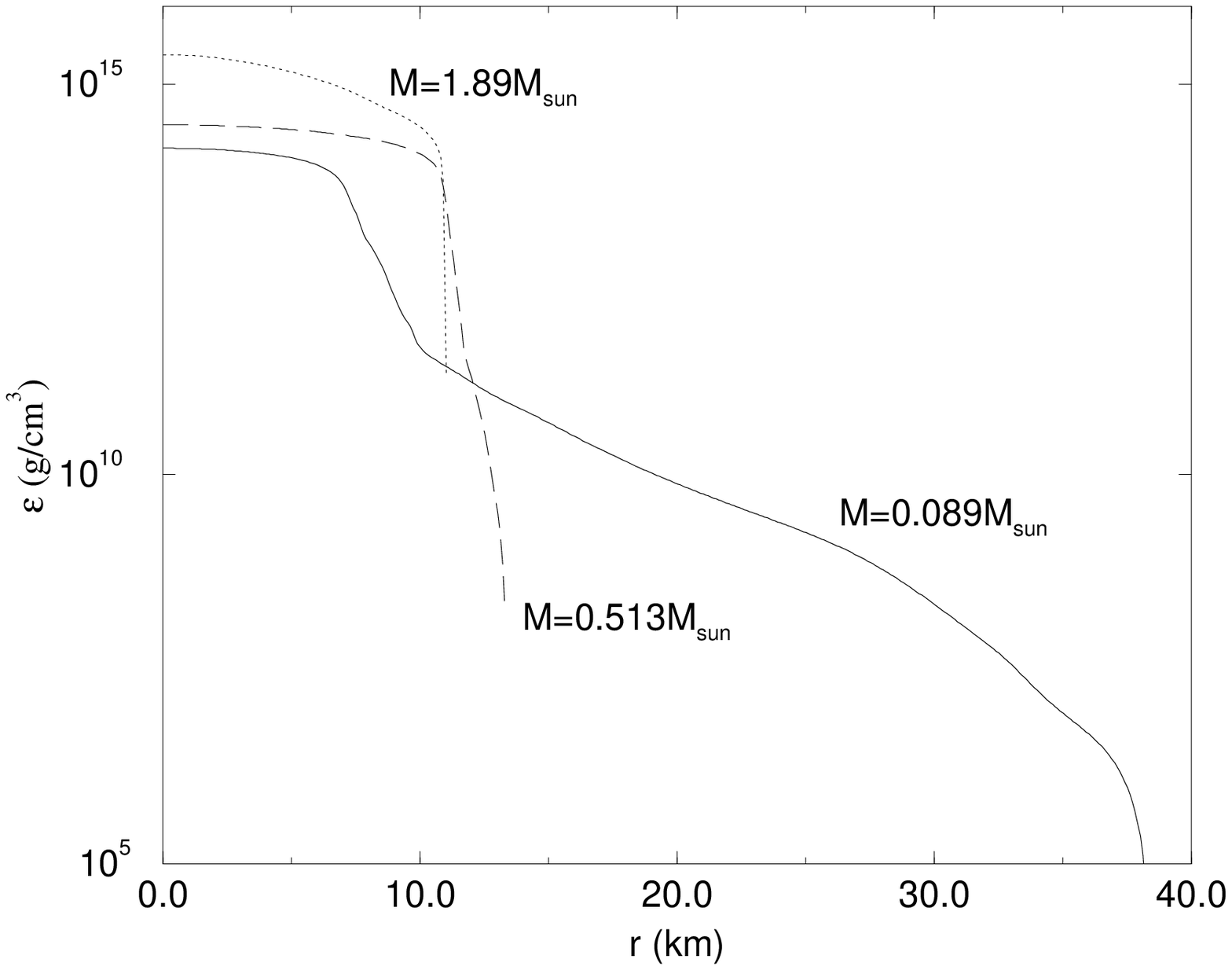}}
\\ {\footnotesize{b)}} \\ \centerline{\epsfysize=6cm \epsfbox{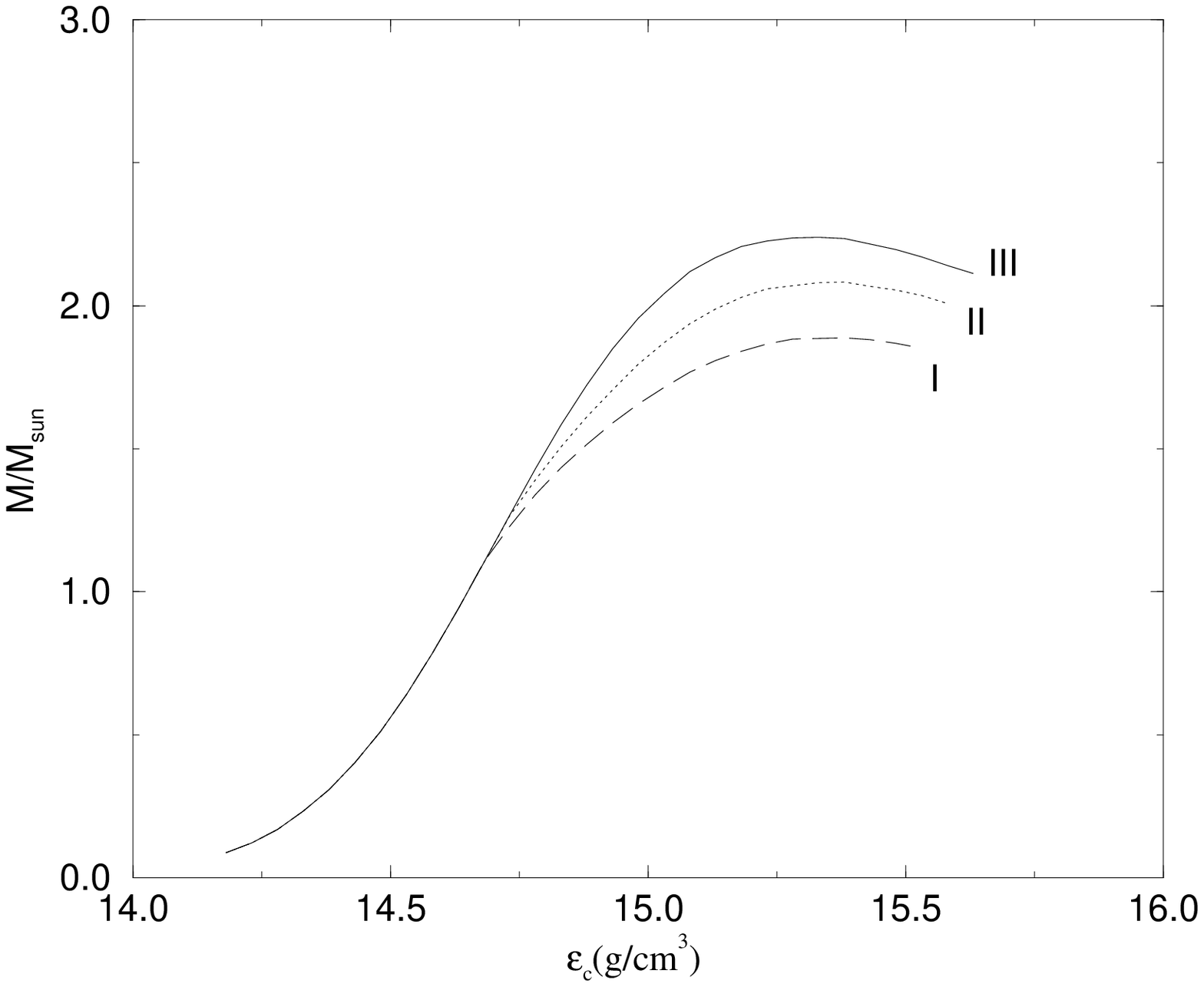}}\\
{\footnotesize{c)}} \\ \centerline{\epsfysize=6cm
\epsfbox{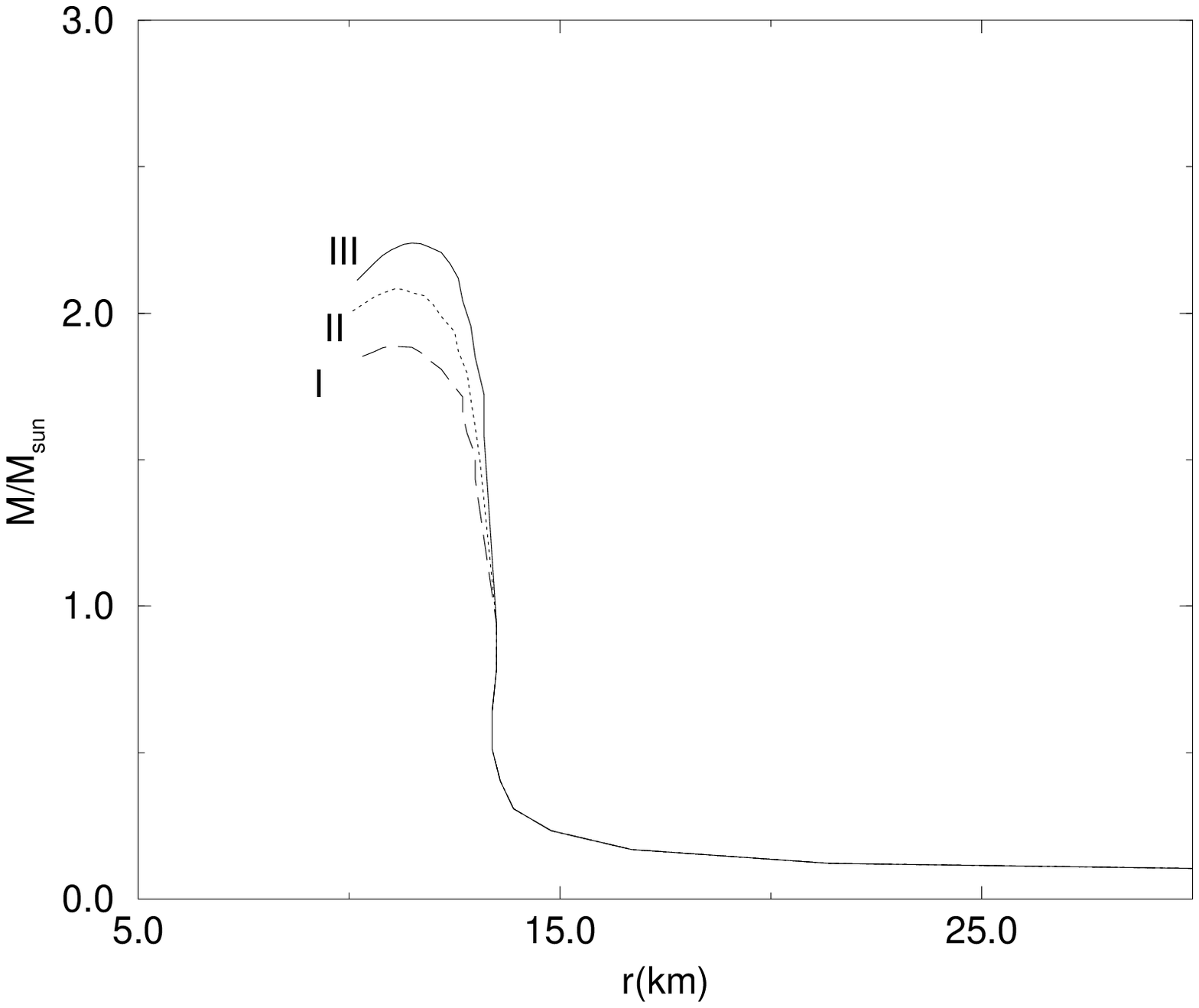}}
\end{tabular}
\end{figure}

To illustrate the importance for neutron star structure of the EOS
at densities greater than $10^{13}\, g/cm^3$, we present in figure
\ref{starnl}.a the results of the energy density as a function of
the radial coordinate for three specific values of the neutron
star mass, for the case without hyperons. We can see that the
energy density is approximately constant for heavier stars and is
found mostly around $10^{14}\,g/cm^3$.

The conversion of nucleons into hyperons reduces the Fermi
pressure associated to the baryons, softening  the equation of
state and lowering, as a consequence, the maximum mass of a
neutron star sequence. In figure \ref{starnl}.b, we see the
neutron star mass and central density relations for the situations
analyzed above. The results indicate, as expected,
 that the presence
of hyperons causes a diminution of the maximum mass of a neutron
star sequence. Typical results for the mass-radius relation are
shown in figure \ref{starnl}.c.

\begin{figure}
\caption{ Radial distribution for matter with (panel {\em a} corresponds to $\chi=\sqrt{2/3}$ and panel
{\em b} to $\chi=1$)
and without (panel {\em c}) hyperons.
\label{radnl}}
\begin{tabular} {c}
{\footnotesize{a)}} \\  \centerline{\epsfysize=6cm
\epsfbox{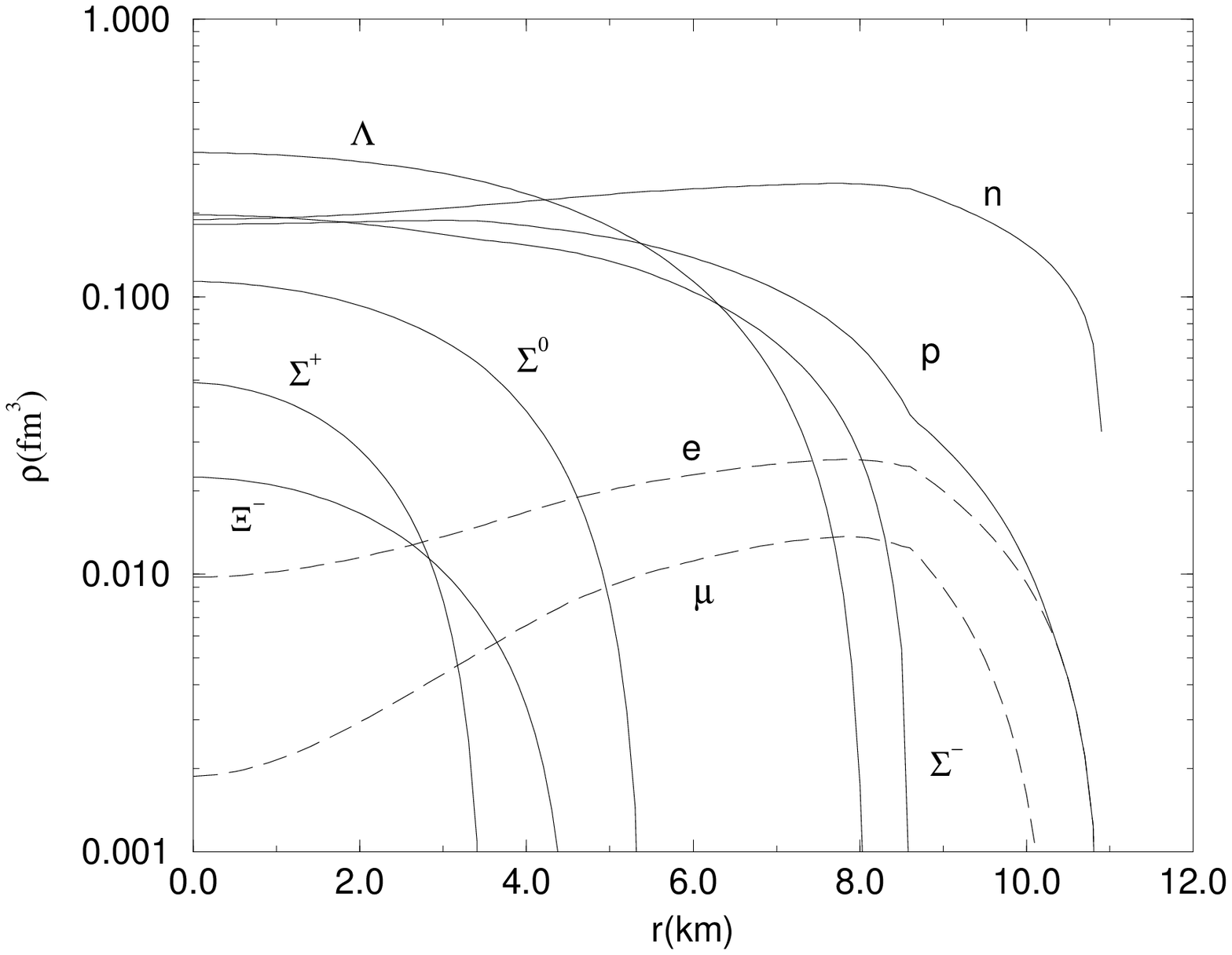}}
\\ {\footnotesize{b)}} \\ \centerline{\epsfysize=6cm \epsfbox{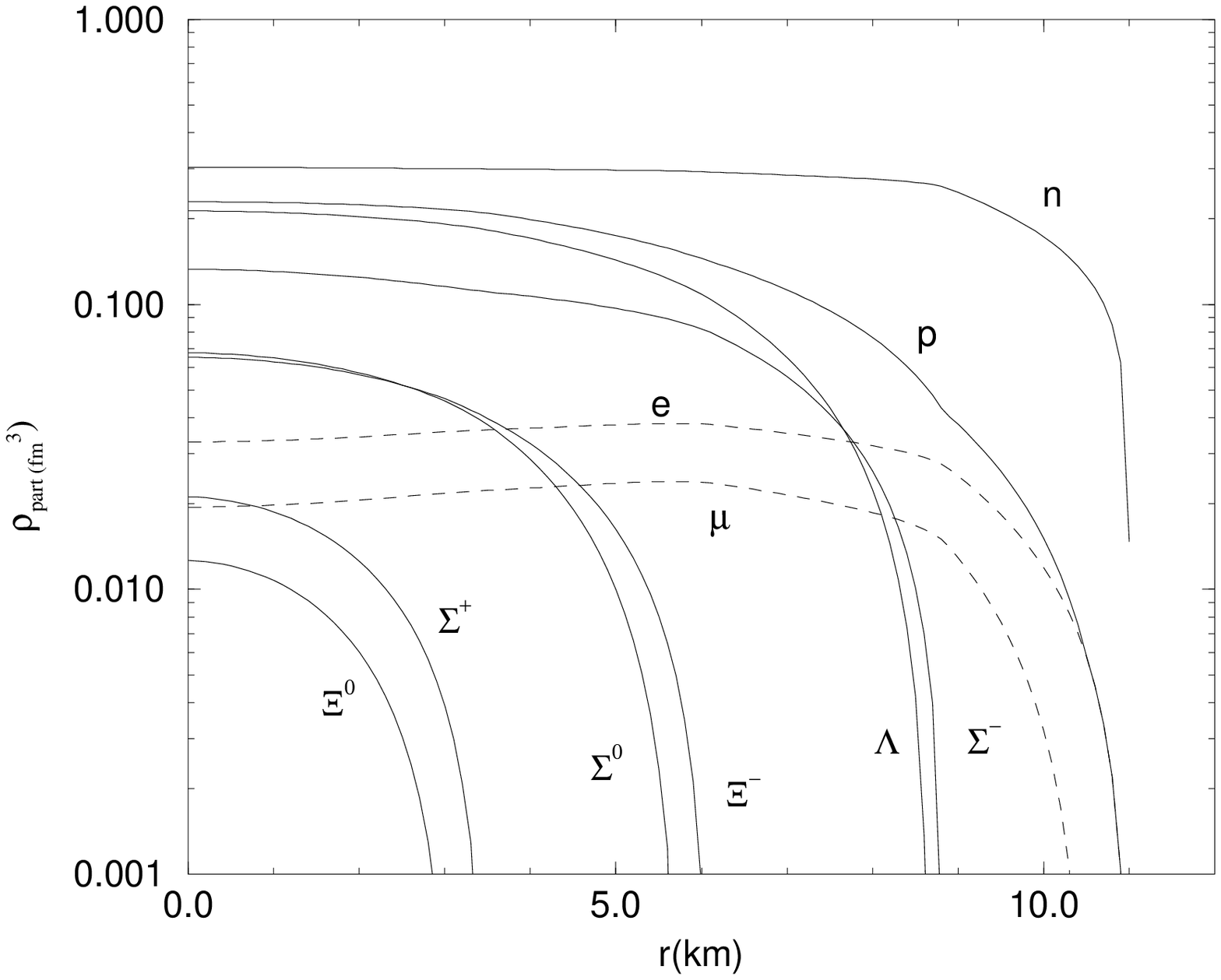}}\\
{\footnotesize{c)}} \\ \centerline{\epsfysize=6cm
\epsfbox{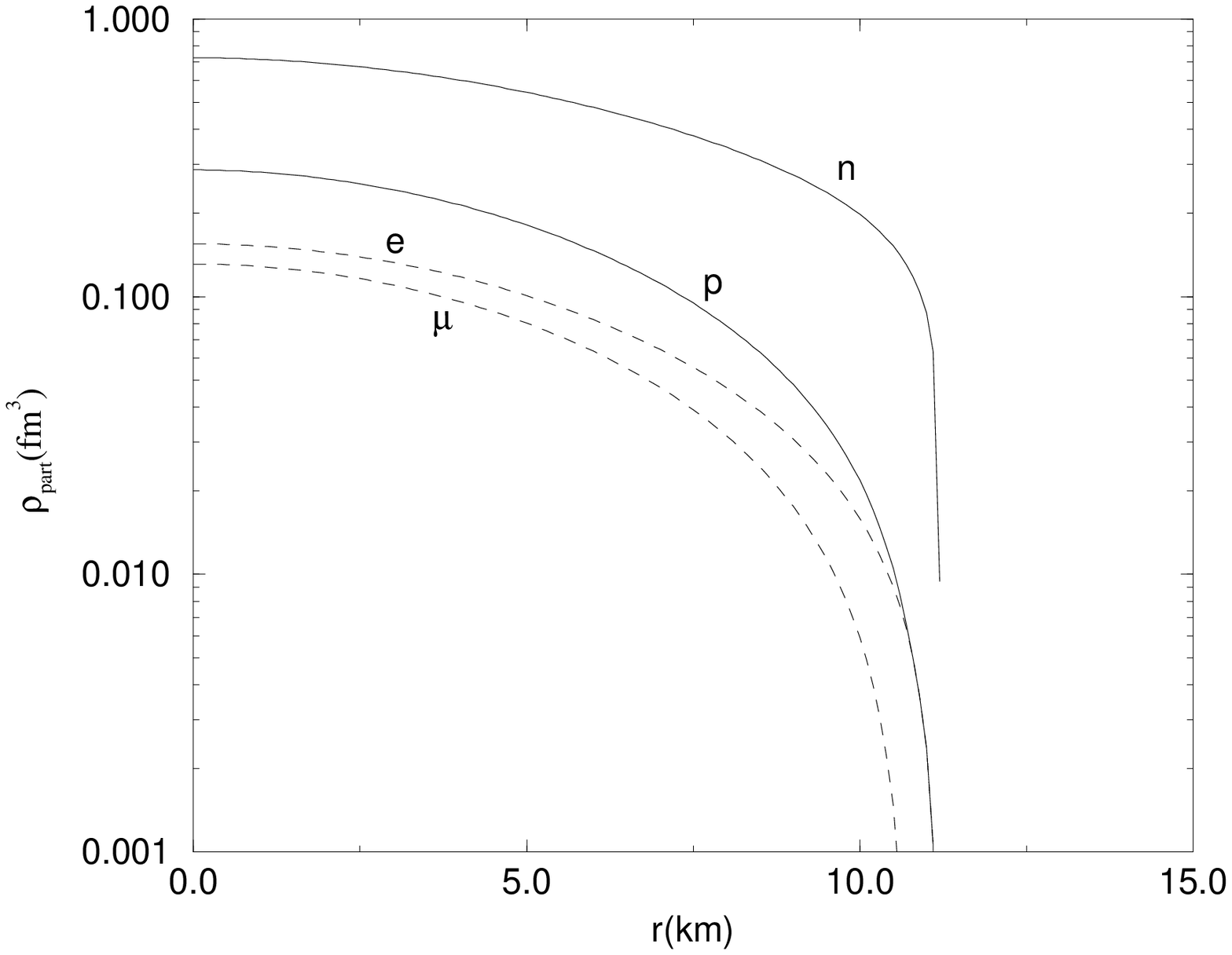}}
\end{tabular}
\end{figure}

Figure \ref{radnl} shows the behaviour of the radial distribution
of baryon and lepton populations. The results indicate that
neutron populations are, in general, dominant. However, in the
case with $\chi=\sqrt{2/3}$ we can see the important contribution
of the $\Lambda$ hyperons in the inner regions of the star.
Moreover,  we have obtained $\sim 11 km$ for the radius of the
star with hyperons and  $\sim 12 km$ without these particles.

\section{Models with Derivative Couplings}
The BB model has two additional coupling constants, $b$ and $c$,
associated to self-interactions of the neutral meson scalar field.
This allows a very good description of two important properties of
nuclear matter, the compression modulus and the nucleon effective
mass, which concerns the high-density behavior of the equation of
state. However, some authors argue that the model suffers of a
very serious problem: the constant c has negative values for
several entries of $M^\star$ and $K$, allowing the energy density
to become unbounded from below for large values of the scalar
meson mean field, leading to unphysical behaviour
\cite{waldhauser}.

In the derivative coupling model, introduced  in 1990
\cite{zimanyi}, the  deficiencies of the original Walecka approach
are eliminated at the cost of making the theory
non-renormalizable. ZM models have been used in the
des\-crip\-tion of sta\-tic pro\-per\-ties of neu\-tron stars
\cite{weber}, $\Delta$ excitations in nuclear matter
\cite{barranco}, bulk properties of finite nuclei
\cite{Chiapparini}, in-medium quark and gluon condensates  and
restoration of chiral symmetry \cite{Dey}, and thermodynamics
properties of nuclear matter \cite{qian,malheiro 98}.

The authors of \cite{zimanyi} have presented two additional
versions of the derivative coupling model. These three models are
known as {\em ZM, ZM2} and {\em ZM3} \cite{delfino,malheiro}.
Concerning the description of static properties of nuclear matter,
the ZM2 model does not exhibit fundamental differences from the ZM
model and will not be considered in the present study.

The Lagrangian density of the ZM and ZM3 models are:
\begin{eqnarray} {\cal{L}}_{_{ZM}} & = & -\bar{\psi}M\psi +
{m^*}^{-1}[\bar{\psi} i \gamma_\mu \partial^\mu\psi -
g_\omega\bar{\psi}\gamma_\mu\psi\omega^\mu] \nonumber \\ & + &
\frac12(\partial_\mu \sigma \partial^\mu \sigma - {m_\sigma}^2
\sigma^2) - \frac14 \omega_{\mu \nu} \omega^{\mu \nu} + \frac12
{m_\omega}^2 \omega_\mu \omega^\mu,  \nonumber \\ \label{lzm}
\end{eqnarray}
and
\begin{eqnarray}
{\cal{L}}_{_{ZM3}} & = & -\bar{\psi}M\psi +
{m^*}^{-1}\bar{\psi} i \gamma_\mu \partial^\mu\psi -
g_\omega\bar{\psi}\gamma_\mu\psi\omega^\mu \nonumber \\
& + & \frac12(\partial_\mu
\sigma \partial^\mu \sigma - {m_\sigma}^2 \sigma^2) -
\frac14 \omega_{\mu \nu} \omega^{\mu \nu} + \frac12 {m_\omega}^2
\omega_\mu \omega^\mu,  \nonumber \\
\label{lzm3}
\end{eqnarray}
where,
\begin{eqnarray}
m^* \equiv (1+\frac{g_\sigma \sigma}{M})^{-1}.
\end{eqnarray}
Expanding $m^*$ in expressions (\ref{lzm}) and (\ref{lzm3}), in
terms of the ratio $\frac{g_\sigma \sigma}{M} < 1$,  we see that
the Walecka and the ZM's models differ essentially on the
replacement of the minimal coupling Yukawa term
$g_\sigma\bar{\psi}\sigma\psi$, by the derivative coupling
contribution $(g_\sigma\sigma/M)
\bar{\psi}\gamma_\mu\partial^\mu\psi$.

 Rescaling in (\ref{lzm}) and (\ref{lzm3}) the nucleon fields in the form
\footnote{In fact, this rescaling introduces a spurious factor
${\cal{F}}=\frac{i}{2}(\bar{\psi}\gamma_\mu\psi)\partial^\mu
ln(m^*(\sigma))$ which does not contribute to the dynamics of the
system \cite{malheiro}.}: \begin{eqnarray} \psi \rightarrow
\sqrt{m^*} \psi \, ,
\end{eqnarray}
the Lagrangian densities of the ZM and ZM3 models may be written
in the general form \footnote{Notice that in the ZM models $m^*M
\equiv M -g_\sigma^\star \sigma$.} \begin{eqnarray} {\cal{L}} & =
& \bar{\psi}\left[ i \gamma_\mu (\partial^\mu + i
g_\omega^\star\omega^\mu) - (M-g_\sigma^\star \sigma)\right] \psi
\nonumber \\ & + & \frac12(\partial_\mu \sigma \partial^\mu \sigma
- {m_\sigma}^2 \sigma^2) - \frac14 \omega_{\mu \nu} \omega^{\mu
\nu} + \frac12 {m_\omega}^2 \omega_\mu \omega^\mu, \nonumber \\
\label{lgeral}
\end{eqnarray}
with:
\begin{equation}
g^\star_\sigma = \frac{g_\sigma}{(1 + \frac{g_\sigma \sigma}{M})}
\, \, ; \, \, \,  g^\star_\omega = \frac{g_\omega}{(1 +
\frac{g_\sigma \sigma}{M})} \, .
\end{equation}
Expression (\ref{lgeral}) reproduces the Lagrangian density of the
Walecka, ZM and ZM3 models with the following replacements:
\begin{eqnarray}
\textrm{Walecka}&:& g_\sigma^\star \rightarrow g_\sigma,
\,\,\,\,\,\,\,\,\,\,\,\,\,\,\, g_\omega^\star \rightarrow
g_\omega; \nonumber
\\
\textrm{ZM} &:& g_\sigma^\star \rightarrow m^* g_\sigma,
\,\,\,\,\,\,\, g_\omega^\star \rightarrow g_\omega;
\label{eq:correspondance}\\ \textrm{ZM3}&:& g_\sigma^\star
\rightarrow m^* g_\sigma, \,\,\,\,\,\,\, g_\omega^\star
\rightarrow m^* g_\omega. \nonumber
\end{eqnarray}

The resulting field equations are: \begin{eqnarray}
[i\gamma_\mu\partial^\mu-(M-g_\sigma^\star \sigma) -
g_\omega^\star \gamma_\mu\omega^\mu]\psi =  0; \\ \nonumber \\
\partial_\nu \omega^{\nu\mu} + {m_\omega}^2 \omega^\mu  =
g_\omega^\star \bar{\psi}\gamma^\mu \psi; \\ \nonumber \\
(\partial_\mu\partial^\mu + {m_\sigma}^2) \sigma  = \frac{\partial
g_\sigma^\star\sigma}{\partial \sigma}\bar{\psi} \psi -
\frac{\partial g_\omega^\star}{\partial \sigma}
\bar{\psi}\omega_\mu \gamma^\nu \psi \,,
\end{eqnarray}
which, after the mean field approximation, become
\begin{equation}
g_\omega \omega_0   =    \frac{g_\omega^2}{m_\omega^2} \rho \, ,
\label{lqe}
\end{equation}
\begin{equation}
 g_\sigma \sigma   =    \frac{g_\sigma^2}{m_\sigma^2}
\left[\frac{1}{g_\sigma}\frac{\partial g_\sigma^\star\sigma}
{\partial \sigma} \rho_s \right]\, , \label{leq}
\end{equation}
for the ZM model and
\begin{equation}
g_\omega \omega_0  =  m^* \frac{g_\omega^2}{m_\omega^2} \rho \, ,
\label{2lqe}
\end{equation}
\begin{equation}
g_\sigma \sigma  = \frac{g_\sigma^2}{m_\sigma^2}\left[
\frac{1}{g_\sigma}\frac{\partial g_\sigma^\star\sigma} {\partial
\sigma} \rho_s - m^* \frac{g_\omega^2}{m_\omega^2}
\frac{1}{g_\sigma g_\omega} \frac{\partial g_\omega \,
^\star}{\partial \sigma} \rho^2 \right] \, , \label{2leq}
\end{equation}
for the ZM3 model. A peculiar aspect of the ZM3 model is  the
coupling between the scalar and vector meson fields, a kind of
coupling  which is not present in other QHD models (Walecka, BB,
ZM). The scalar and vector mean field potentials are defined to be
\begin{equation}
S \equiv - g_\sigma^\star \sigma \, , \label{s}
\end{equation}
and
\begin{equation}
V \equiv g_\omega^\star \omega_0 \, . \label{v}
\end{equation}
In these models, the  expressions for the energy density and
pressure  are equivalent to the corresponding expressions
(\ref{3nldener}) and (\ref{3nlpressao}) of the non-linear model,
with $b=c=0$ and the replacements shown in
(\ref{eq:correspondance}).

The values of  the scalar $(g_\sigma/m_\sigma)^2$ and vector
$(g_\omega/m_\omega)^2$ coupling constants which reproduce the
saturation density, $\rho_0 = 0.15 fm^{-3}$, and the saturation
energy, $B = -16 MeV$, in case of the ZM and ZM3 models are shown
in  table \ref{tabelzm}; the table also contains the values of the
nucleon effective mass, the compression modulus of nuclear matter,
the scalar and vector mean fields S and V for both models, and
values for the isovector coupling constant
$(g_\varrho/m_\varrho)^2$ which reproduce the symmetry energy
coefficient $a_4 = 32.5$ MeV.

\begin{figure}
\caption{Equation of state ({\em panel a}) and nucleon effective
mass ({\em panel b}) for the models W, BB, ZM and ZM3.
\label{figzm1b}}
\begin{tabular} {c}
{\footnotesize{a)}} \\  \centerline{\epsfysize=6cm
\epsfbox{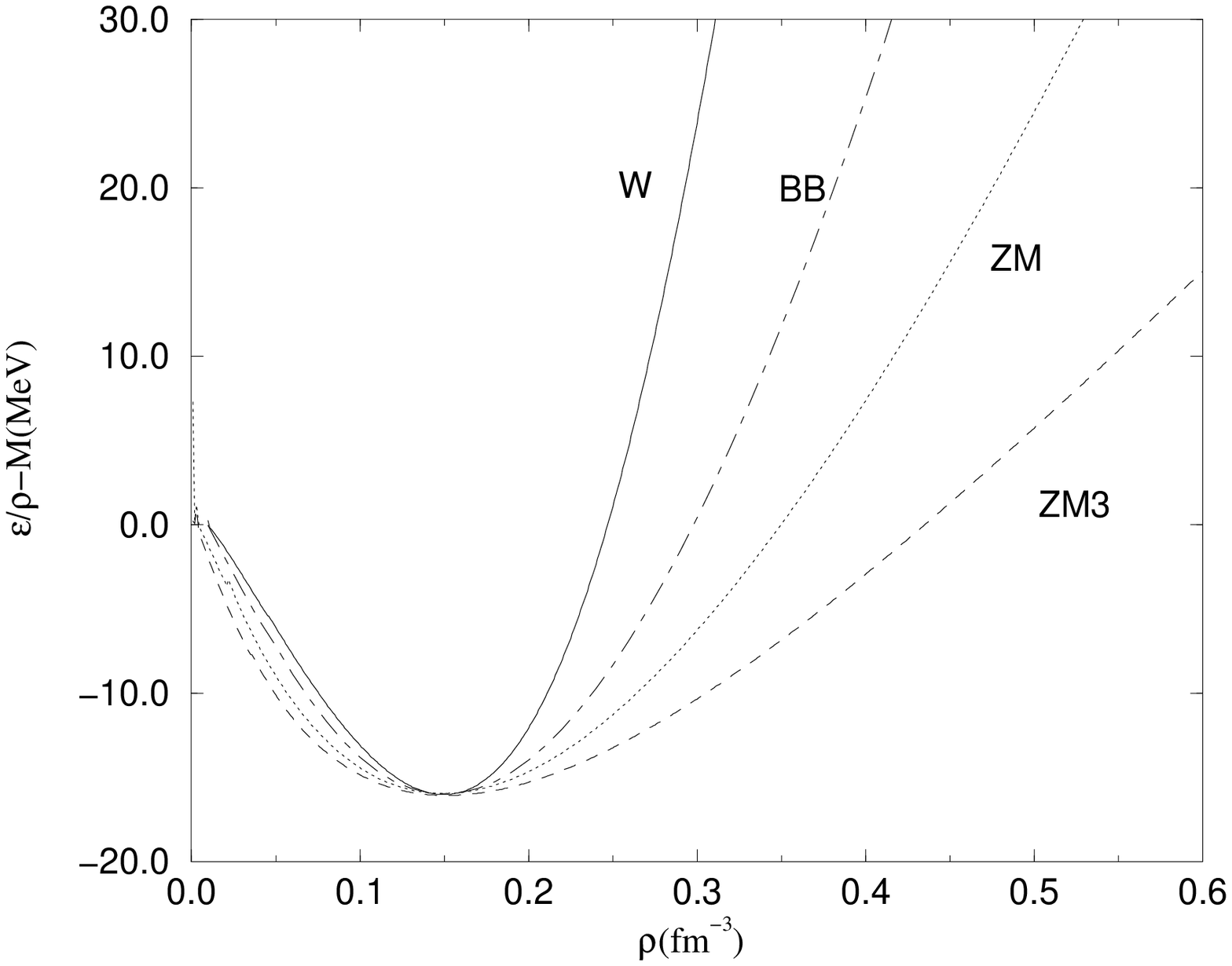}}
\\ {\footnotesize{b)}} \\ \centerline{\epsfysize=6cm \epsfbox{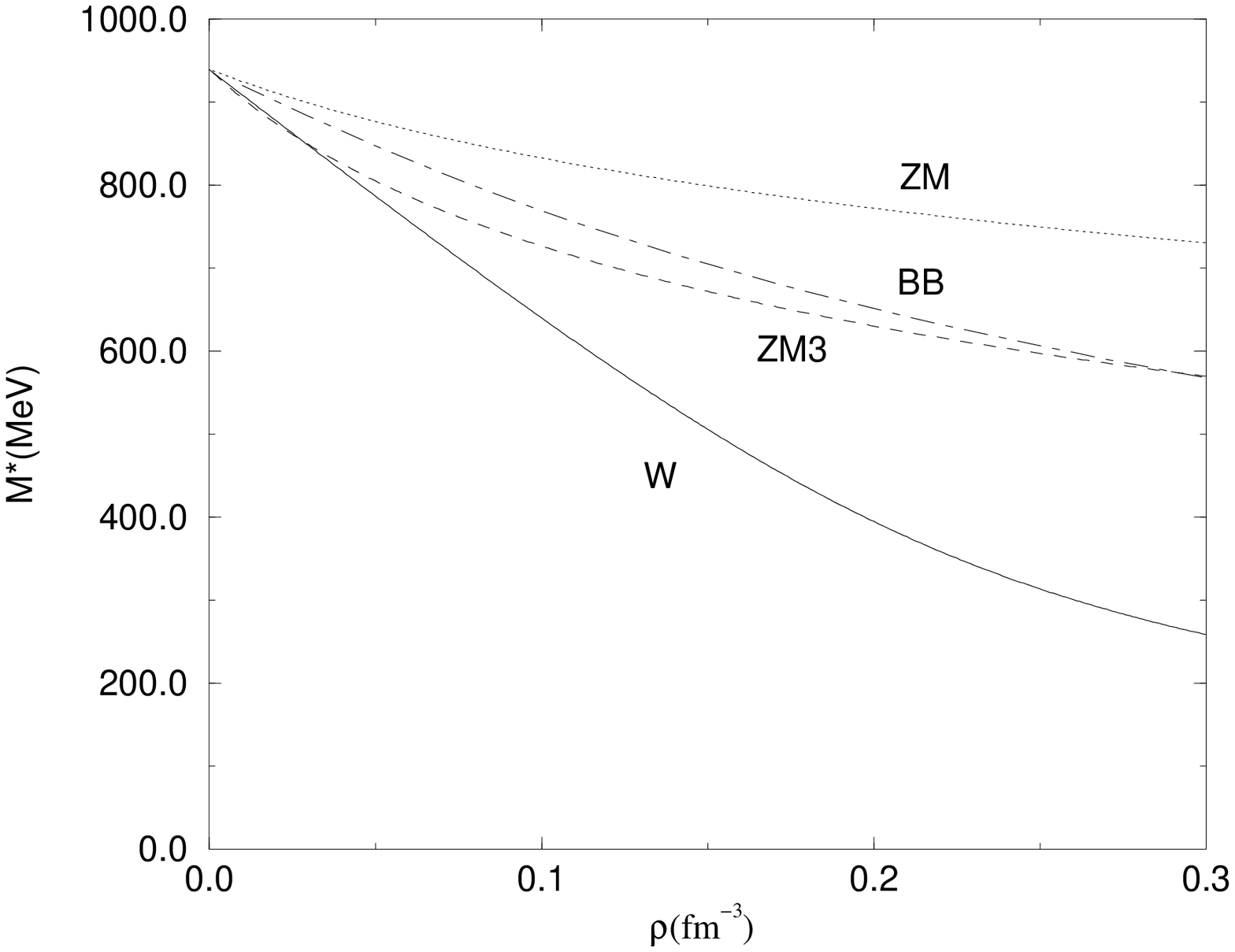}}
\label{figzm1a}
\end{tabular}
\end{figure}

\begin{figure}
\caption{Evolution of the coupling constants $g^\star_{\sigma,\omega}/m_{\sigma,\omega}$ with increasing
density for ZM model ({\em panel a}) and ZM3 model ({\em panel b}). \label{figzm2a}}
\begin{tabular} {c}
{\footnotesize{a)}} \\  \centerline{\epsfysize=6cm
\epsfbox{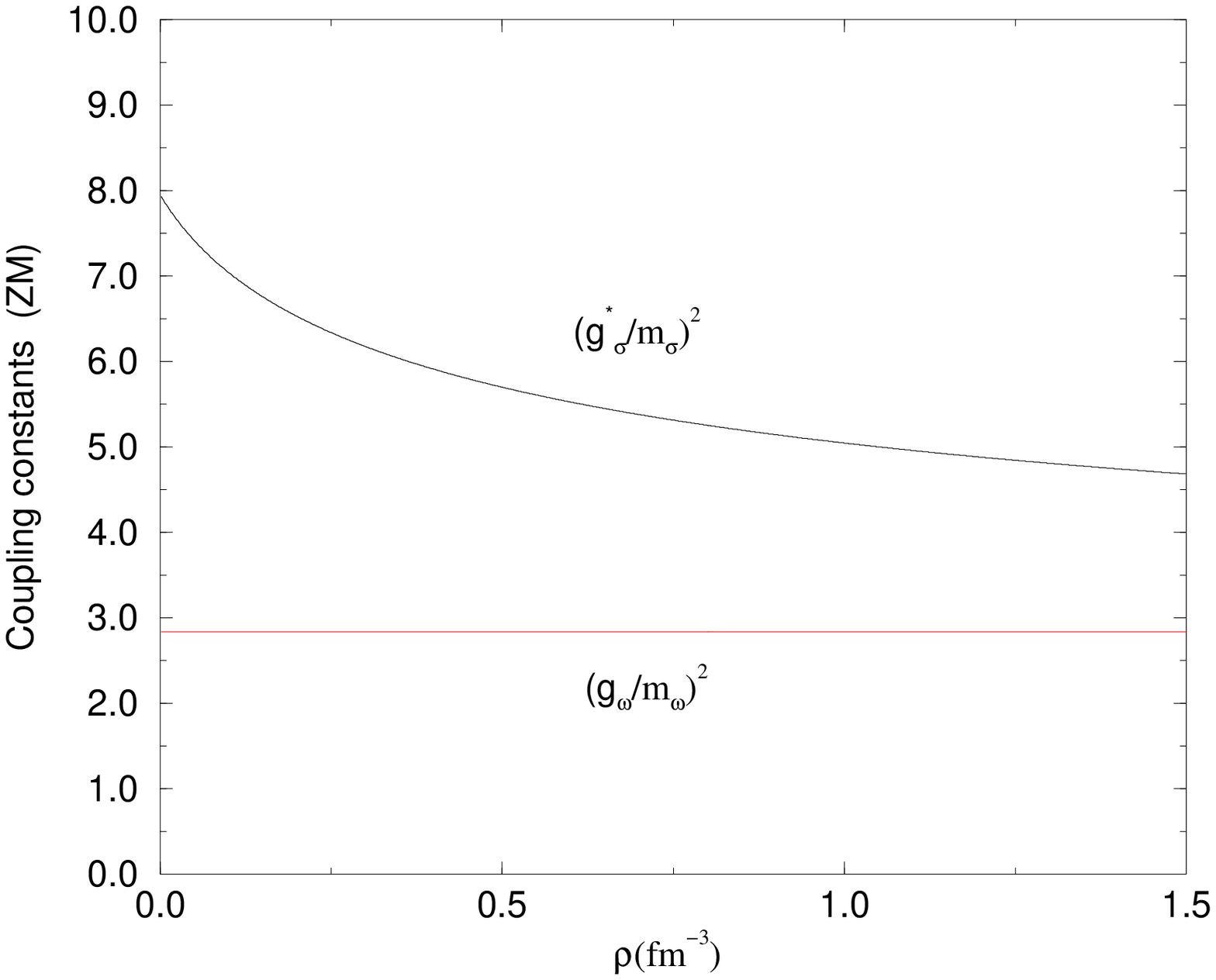}}
\\ {\footnotesize{b)}} \\ \centerline{\epsfysize=6cm \epsfbox{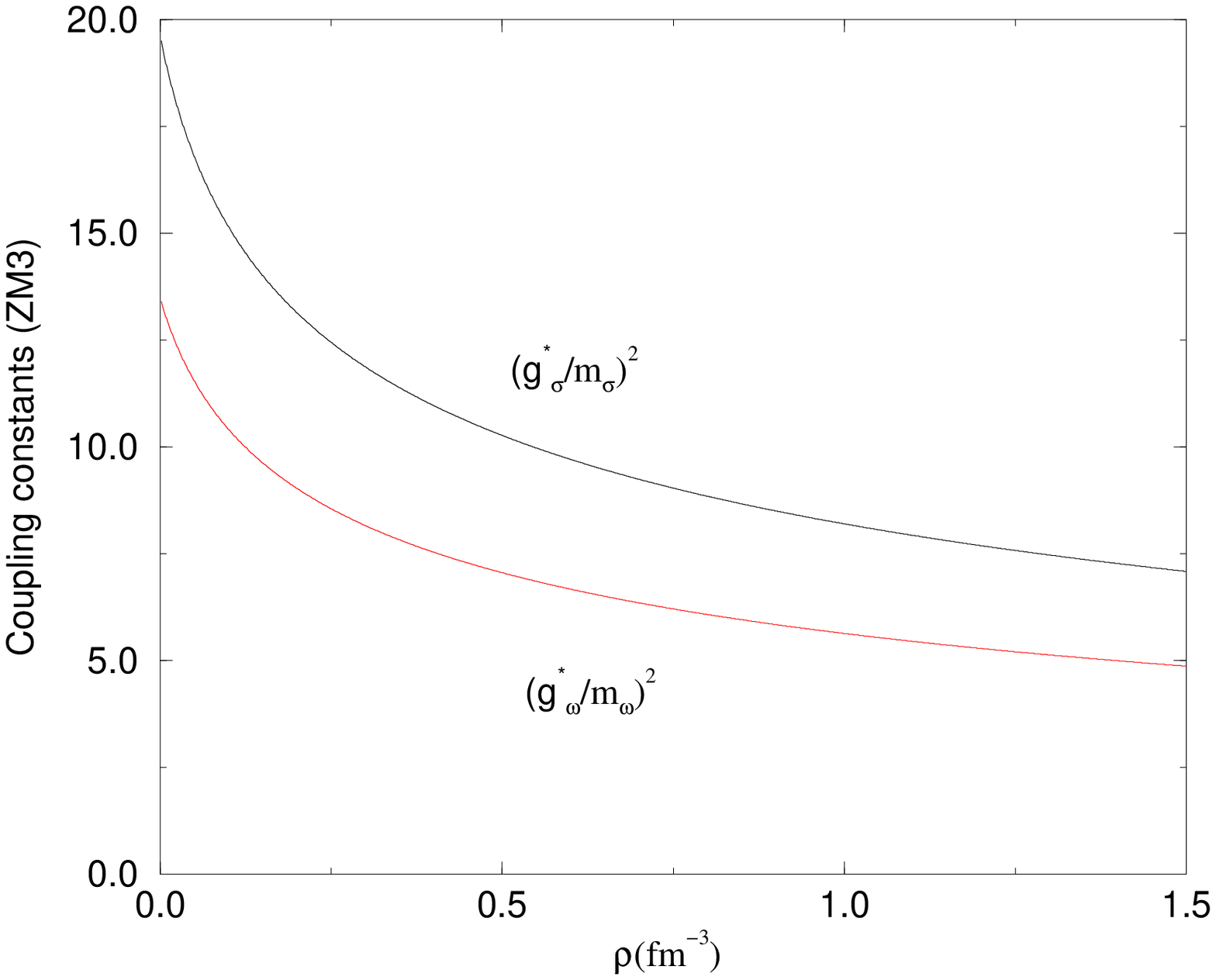}}
\end{tabular}
\end{figure}

From figure  \ref{figzm1a}.a we see that the ZM and  ZM3 models
produce a softer  equation of state when compared to the
corresponding results for the Walecka and BB models. The ZM and
ZM3 models also produce higher values for the nucleon effective
mass as shown in figure \ref{figzm1a}.b.
Figures \ref{figzm2a}.a and \ref{figzm2a}.b show the behavior of
the non-linear coupling constants $({g_\sigma}^\star/m_\sigma)^2$
and $({g_\omega}^\star/m_\omega)^2$ as a function of density for
the ZM and ZM3 models. Notice that the {\em coupling strength
decreases with increasing density}.

A interesting comparison among the models considered in this
section (Walecka, BB, ZM and ZM3) is related to their
corresponding values for the relativistic coefficient defined as:
\begin{eqnarray} R \equiv \frac{\rho_s}{\rho} = \int_0^{k_{F}} d^3
k \frac{M^*}{\sqrt{k^2 + M^{* 2}}} /\int_0^{k_{F}} d^3 k \, .
\label{coefrel}
\end{eqnarray}
According to this definition, for a {\em less relativistic} model
$\rho_s$ becomes closer to $\rho$ because $k \ll M$. For the
models discussed in this section we have
$
R_{W}  =  0.931, \, R_{BB}  =  0.963 ,\, R_{ZM}  =  0.970, \,
R_{ZM3}  =  0.959.
$
The results of the
ZM3 model for the relativistic coefficient
are closer to the corresponding results
of  the Walecka model.

After the work of Zimanyi and Moszkowski, many authors started to
explore extensively  the freedom in the choice of the meson-baryon
interaction. As explained below, based on these works we introduce
in this paper a new class of models which enable us to make direct
comparisons among the properties of nuclear matter and neutron
stars. We also employ special cases of this class, namely Walecka,
ZM and ZM3 models, in the description of static properties of
these stellar objects.

\section{A Class of Non-Linear Relativistic Models}
In this section we  propose a new class of relativistic hadronic
models which exhibits a non-linear parameterization of the
intensity of the meson couplings and incorporate some QHD models
found in the literature.  We study, through this comprehensive
approach, the influence of non-linear meson-nucleon couplings in
the nucleon effective mass, compression modulus of nuclear matter,
relativistic coefficient  and static properties of neutron stars.

\subsection{Phenomenological Lagrangians}
Koepf {\em et al.} \cite{koepf} have studied the contribution of
the term ${\cal{L}}_{\sigma N} = M \bar{m} ^\star (\sigma)
\bar{\psi}\psi$ using the phenomenological couplings shown in
table \ref{tabelfe}. The first term in the table corresponds to
the Walecka  model and the last term to the ZM model. Glendenning
{\em et al.} \cite{weber} have analyzed a coupling term of the
type \begin{eqnarray} \bar{m}^\star = (1- \frac{g_\sigma \sigma}{2
M})(1+ \frac{g_\sigma\sigma}{2M})^{-1},
\end{eqnarray}
obtaining $M^*/M = 0.796$ and  $K = 265 MeV$.

In these studies the authors have assumed, at first order in
$g_\sigma \sigma/M$, a similar expression for the nucleon
effective mass as introduced by Walecka: $M^*/M \sim 1 - g_\sigma
\sigma / M$. This basic assumption is essential, since $g_\sigma
\sigma$ represents the nucleon mass shift in the scalar meson mean
field condensate. The experimental results suggest, at nuclear
saturation density,
 $M^*/M\sim0.70$, giving
$g_\sigma\sigma/M\sim0.3$. Thus, these different models found in
the literature just add  scalar self-coupling corrections terms to
the corresponding expression of the Walecka model.

On the basis of the various approaches found in the literature, we
propose a new phenomenological Lagrangian with a non-linear
parameterization, through mathematical constraints
($\lambda$,$\beta$ and $\gamma$ parameters), of the analytical
form of the meson-baryon couplings: 
\begin{eqnarray} {\cal{L}} & = &
 \sum\limits_{B} \bar{\psi}_B [i \gamma_\mu \partial^\mu - (M_B-
g_{\sigma B}^\star \sigma)-g_{\omega  B}^\star \gamma_\mu
\omega^\mu ]\psi_B \nonumber \\ & - & \sum\limits_{B} \psi_B
[\frac12 g_{\varrho B}^\star \gamma_\mu
\mbox{\boldmath$\tau$}\cdot \mbox{\boldmath$\varrho$}^\mu] \psi_B
 +  \sum\limits_\lambda \bar{\psi}_\lambda [i \gamma_\mu \partial^\mu -
m_\lambda] \psi_\lambda \nonumber \\ & + & \frac12(\partial_\mu
\sigma \partial^\mu \sigma - {m_\sigma}^2 \sigma^2) - \frac14
\omega_{\mu \nu} \omega^{\mu \nu} + \frac12 {m_\omega}^2
\omega_\mu \omega^\mu \nonumber \\ & - & \frac14
\mbox{\boldmath$\varrho$}_{\mu \nu} \cdot
\mbox{\boldmath$\varrho$}^{\mu \nu} + \frac12m_\varrho^2
\mbox{\boldmath$\varrho$}_\mu \cdot \mbox{\boldmath$\varrho$}^\mu
\nonumber \\
\end{eqnarray}
where
\begin{equation}
g_{\sigma B}^\star \equiv m^\star_{\lambda B} g_{\sigma}  \, \, ;
\, \, g_{\omega B}^\star \equiv m^\star_{\beta B} g_{\omega}  \,
\, ; \, \, g_{\varrho B}^\star = m^\star_{\gamma B} g_{\varrho} \,
\end{equation}
and
\begin{equation}
m^\star_{n B} \equiv (1 + \frac{g_{\sigma} \sigma}{n M_B})^{-n} \,
; \, \, n = \lambda, \beta, \gamma \, \label{gefome1}.
\end{equation}
In these expressions, we assume $\lambda$, $\beta$ and $\gamma$ as
real and positive numbers, since this is the range of best
phenomenology. As discussed in the previous section, essentially
what was done is the introduction of a rescaling of the scalar and
vector coupling terms of the Walecka model. For instance, in case
of the scalar contribution we have made the replacement
\begin{eqnarray} g_\sigma\sigma\bar{\psi}\psi \rightarrow
g^{\star}_{\sigma} \bar{\psi} \sigma \psi =
\frac{g_\sigma\sigma}{\left(1+\frac{g_\sigma\sigma}{\lambda M}
\right)^\lambda} \bar{\psi}\psi  \, .
\end{eqnarray}
Similar interaction terms may be associated to the vector and
isovector sector of the Lagrangian density. Notice that we have
assumed an universal coupling by setting $g_{(\sigma,\omega,\varrho)B}\to
g_{(\sigma,\omega,\varrho)}$.

Table \ref{corres} exhibits the correspondence between this and
the other models discussed in this work with specific values of
$\lambda$, $\beta$ and $\gamma$. One of the main intentions  of
the present study is to consider values of these parameters which
give better results for nuclear matter and neutron star properties
when compared to the corresponding results of the traditional
models discussed in this work. As far as we know, the first
extensions of the ZM-like models to applications to neutron star
matter with the inclusion of hyperons ($\Lambda,\Sigma^-$) and
leptons was done in \cite{Bhattacharyya98}. Here we consider these
known models as well as intermediate values of the parameters of
our non-linear coupling to obtain results for neutron star
properties and relate them to nuclear matter saturation
observables.

Using the mean field approximation, the field equations in our
approach become
\begin{eqnarray}
[i\gamma_\mu\partial^\mu&-&g_{\omega B}^\star \gamma_0 \omega^0 -
(M_B-g_{\sigma B}^\star\sigma)]\psi_B = 0 \, ; \\
 g_\omega \omega_0 & = &
\left(\frac{g_\omega}{m_\omega}\right)^2 \sum\limits_B m^*_{\beta
B}  \rho_B; \\ g_\varrho \varrho_{03} & = &
\left(\frac{g_\varrho}{m_\varrho}\right)^2 \sum\limits_B
m^*_{\gamma B}  I_{3B} \rho_B; \\ - g_\sigma \sigma & = &
\left(\frac{g_\sigma}{m_\sigma}\right)^2 \left[ \sum\limits_B
\left(\frac{F_B(\sigma)}{g_\sigma}\right) \varrho_{s,B} \right.
\nonumber
\\ & + & g_\omega\omega_0\left(\frac{g_\sigma}{m_\sigma}\right)^2
\sum\limits_B\left(\frac{G_B(\sigma)}{g_\sigma g_\omega}\right)
\rho_B \nonumber \\ &+& \left.
g_\varrho\varrho_{03}\left(\frac{g_\sigma}{m_\sigma}\right)^2
\sum\limits_B\left(\frac{H_B(\sigma)}{g_\sigma g_\varrho}\right)
I_{3B} \rho_B \right] \, ,
\end{eqnarray}
where $F_B(\sigma)$, $G_B(\sigma)$ and $H_B(\sigma)$ are given by
\begin{equation} \frac{F(\sigma)}{g_\sigma} = - m^*_\lambda +
\frac{g_\sigma\sigma}{M} ({m^*_\lambda})^{(\lambda+1)/\lambda} \,
, \label{fg1}
\end{equation}
\begin{equation}
\frac{G(\sigma)}{g_\sigma g_\omega} = -\frac{(m^*_\beta)
^{(\beta+1)/\beta}}{M} \,,\label{fg2}
\end{equation}
and
\begin{eqnarray}
\frac{H_B(\sigma)}{g_\sigma g_\varrho} = -\frac{(m^*_\gamma)
^{(\gamma+1)/\gamma}}{M_B} \, .
\end{eqnarray}

From the eigenvalues of the Dirac equation, the Fermi energy is:
\begin{eqnarray} \mu_B(k) &=&g_{\omega  B}^\star
\omega_0 \!+\! g_{\varrho B}^\star
\varrho_{03}I_{3B} \! +\!  \sqrt{k_{F,B}^2\! +\! (M_B\!
-\!g_{\sigma B}^\star\sigma)^2}. \nonumber \\
&&\label{potlamb}
\end{eqnarray}
The expressions for the scalar and vector potentials (S) and (V)
are \begin{eqnarray} S  =   -m^*_\lambda g_\sigma \sigma \,
,\,\,\, V  =  m^*_\beta g_\omega \omega_0 \, .
\end{eqnarray}
We can see that this model allows some control on the intensity of
the scalar and vector mesons mean-field potentials. For instance,
to the variations of $\lambda$ between $0$ and $1$, keeping $\beta
= \gamma = 0$, we obtain
 values
of $S, \,V,\, M^*$ and $K$ which correspond to  the intermediate
region of values of Walecka and  ZM models. Similarly, for values
of $\lambda$, $\beta$ and $\gamma$ between $0$ and $1$, we can
find intermediate results  between the Walecka and the ZM3 models.

Indeed, the range of possible values for the parameters of the
theory is not very large. Due to the form of the general coupling
terms (see equation (\ref{gefome1})), there occurs a rapid
convergence to an exponential form. Taking $\lambda$, $\beta$, and
$\gamma$  $\to \infty$, we have \footnote{Notice that this form of
the exponential coupling term is different from the corresponding
one of model 3 shown in table \ref{tabelfe}.}:
\begin{eqnarray}
g_\sigma^\star & \rightarrow & e^{\frac{-g_\sigma\sigma}{M}} g_\sigma \, ;
\label{expo1}  \, \, \,
g_\omega^\star \rightarrow e^{\frac{-g_\sigma\sigma}{M}} g_\omega \, ;
\label{expo2}
\\ \nonumber
g_\varrho^\star & \rightarrow & e^{\frac{-g_\sigma\sigma}{M}} g_\varrho \, .
\label{expo3}
\end{eqnarray}
As we will see later, for $\lambda$ and/or $\beta=\gamma > 2$  the
results of this model do not strongly differ from the results of
the model with exponential coupling.

In this work, we shall  consider two cases:
\begin{itemize}
\item Scalar (case S): we consider in this case
variations of $\lambda$
with $\beta = \gamma = 0$;
this case contains the results of the
Walecka and  ZM models.
\item Scalar-Vector (case S-V):
we consider in this case variations of $\lambda$, with $\beta =
\gamma  = \lambda$; Walecka and ZM3 models belong to this
category.
\end{itemize}
Notice that the models we are considering may be uniquely
specified by the $\lambda$ parameter. Walecka model belongs to
both categories because in this model the mathematical parameters
$\lambda$, $\beta$ and $\gamma$ are null; but it does not present
scalar-vector interaction contributions. Taking the limit
$\lambda\rightarrow\infty$ for the S case, we have a model with
exponential coupling; in the figures we refer to this asymptotic
model as EXP/S. Similarly, for
$\lambda,\beta,\gamma\rightarrow\infty$ we have the asymptotic
model EXP/S-V. In the following we exploit the nuclear properties
of our non-linear class of models.

\subsection{Nuclear Properties}
We de\-ter\-mi\-na\-te the cou\-pling cons\-tan\-ts
$g_{\sigma,\omega,\varrho}/m_{\sigma,\omega,\varrho}$
in this mo\-del by  fo\-llow\-ing the sa\-me pro\-ce\-du\-re pre\-sen\-ted in
sec\-tion \ref{3secpop}. For each ca\-se, S and S-V, we obtain
numerical values for $g_\sigma/m_\sigma$, $g_\omega/m_\omega$ and
$M^\star$, as a function of $\lambda$. Thus, we can also determine
$g_\varrho/m_\varrho$. We also find an analytical expression for
the compression modulus of nuclear matter as a function of the
nucleon effective mass (see Appendix \ref{appendixb}).

Figure \ref{figcon1} exhibits the dependence of the coupling
constants on the $\lambda$ parameter; it is interesting to note
the regular relative behaviour of the coupling constants
$(g_\sigma/m_\sigma)^2$ and $(g_\omega/m_\omega)^2$. The results
indicate that the scalar model suffers a $\lambda$ dependent
saturation process in a small range of values for this parameter.
On the other hand, the results also reveal that the scalar-vector
model exhibits a wider range of values of $\lambda$.

In figure \ref{figcon2} we show results corresponding to the $S_0
\times V_0$ plane for different values of $\lambda$. To understand
these results one should consider expression (\ref{potlamb}) for
symmetric nuclear matter (with $k_n=k_p$ and the contributions of
hyperons and leptons turned off) at saturation density:
\begin{eqnarray} \mu_0 = V_0 +
\sqrt{k_0^2+(M+S_0)^2} \, . \label{mu0}
\end{eqnarray}
Since $\mu_0 = \varepsilon_0/\rho_0 = M + B$ and using $k_0/M$ and
$S_0/M$ as expansion parameters,
 equation
(\ref{mu0}) may be approximately expressed in the form
\begin{eqnarray} B \sim (V_0+S_0) + (1 - S_0/M) \frac{k_0^2}{2M}
\, . \label{linear}
\end{eqnarray}
This is a very well known and interesting result: the Lorentz
structure of the interaction leads to a new energy scale and the
small nuclear binding energy per nucleon arises from a delicate
cancellation between the large components of the scalar attraction
and vector repulsion plus an additional kinetic energy term of a
nucleon with non-linear mass $M+S_0$ ($ \sim M/(1 - S_0/M)$) at
the Fermi level. Expression (\ref{linear}) leads to a linear
relation between $S_0$ and $V_0$, explaining the results of figure
\ref{figcon2}.

Results for the compression modulus as a function of the $\lambda$
parameter for the S and S-V cases are shown in figure
\ref{4kmlam1}; according to these results, there is a minimum of
$K$, in the S case, for $\lambda \sim 0.2$. In figure
\ref{4kmlam2} we show the monotonic behaviour of the nucleon
effective mass, at saturation density, as a function of $\lambda$.

Figure \ref{4km1} presents the relation between the compression
modulus and the ratios of the nucleon effective mass, $K\times
M^*/M$, for the S and S-V cases. From the results, one can see
that to higher values of the nucleon effective mass it corresponds
lower values of the compression modulus: to understand this
behaviour one should remember that the repulsive mean-field vector
potential $V_0$ is proportional to $M - M^*$ (see figure
\ref{figcon2}) . From the results found in the literature
\cite{Furnstahl96}, the nucleon effective mass and the compression
modulus should be in the range $0.6 < M^*/M < 0.7$ and $200 < K <
300 $ MeV. As stressed before, the scalar case leads to a narrow
interval of values of $\lambda$ while the scalar-vector case has a
broader range. {\it Accordingly, case S gives reasonable results
in the range $\lambda \sim 0.05 - 0.07$ and case S-V better
results for $\lambda \sim 0.16 - 0.4$. The analysis also reveals
the strong similarity between the results of ZM-like models and
those with exponential couplings.}

The results of figure \ref{4km2} show that models with higher
relativistic coefficients (W and ZM3) have a smaller scalar
potential $S$. To understand these results, one should consider
the definition of R (equation \ref{coefrel}). For less
relativistic models, $R \to 1$, since for these models  $\rho_s$
becomes closer to $\rho$ because $k/M^* \ll 1$.

We turn now to the high density behaviour of these approach by
considering its application to the neutron star environment.

\subsection{Neutron Star Properties}
In this section we consider the determination of neutron stars
properties using our class of non-linear models with the inclusion
of hyperon and  lepton  degrees of freedom.

Following the same procedure of section \ref{3secpop}, we solve a
system of transcendental equations taking into account chemical
equilibrium, baryon number  and electric charge conservation and
the equations for the $\sigma$-field. We then obtain the EOS for
our system. The resulting expressions for the energy density and
pressure are, again, similar to (\ref{3nldener}) and
(\ref{3nlpressao}) but with $b=c=0$ and
$M^\star_B=M_B-g^\star_{\sigma B}\sigma$. Combining this EOS  with
the TOV equations we obtain values for static properties of
neutron stars (mass, radius, baryon composition among others) as
functions of the central density.

We pre\-sen\-t ex\-pli\-ci\-tly he\-re the neu\-tron star
phe\-no\-me\-no\-lo\-gy for the Wa\-le\-cka and
Zi\-ma\-nyi-Mosz\-ko\-wski mo\-dels, namely the original ZM and
also the variant ZM3 model, whose results are presented here for
the first time.

The predictions for neutron star masses as a function of the
central density in Walecka, ZM and ZM3 models are shown in figure
\ref{4zmotv2}. ZM model predicts a maximum mass of approximately
$1.6 M_\odot$, in the limit of acceptability  for the mass of a
pulsar. In particular, the ZM3 model is very soft and predicts a
very small maximum neutron star mass, $\sim0.72 M_\odot$. It may
be surprising, at a first glance,  that the maximum neutron star
mass for the Walecka model with hyperons ($2.77M_\odot$) exceeds
the well known result ($2.6M_\odot$) found in \cite{walecka 86}
for stars just composed of neutrons, since the addition of other
particles softs the equation of state, lowering the resulting star
mass. This apparent contradiction can be explained by the  extreme
sensibility of this kind of theory on the specific choice of the
values of the binding energy and saturation density. The authors
cited above have used $B=-15.75\,\,MeV$ and
$\rho_0=0.19\,fm^{-3}$; with this choice we were able to reproduce
their results. However, with our choice for these quantities,
which is widely used in the recent literature, we get for the mass
of a star composed only by neutrons the value of $3.05M_\odot$,
that is, a difference of almost a half solar mass! Using the
constants from \cite{walecka 86} ($a_4=33.6\,MeV$), we get
$2.33M_\odot$ for the mass of a neutron star with the inclusion of
hyperons and leptons. In this way, extrapolation for neutron star
densities from the fitting of $B$ and $\rho_0$ at saturation needs
more precision on the choice of the values for these quantities.

Figure \ref{4zmpop2}  shows the behaviour of the chemical
potentials and field intensities in the Walecka, ZM  and ZM3
models.  These results should be compared to the corresponding one
obtained with the BB model. We observe the same saturation of the
electron chemical potential at $\sim 200\,MeV$ for the ZM and BB
models with universal coupling (see figure \ref{3Ipop1}.d). ZM3
model presents higher values for $-g_\varrho\varrho_{03}$, as
compared to the other models; this behaviour is connected with the
large value for $g_\varrho/m_\varrho$ in this model ({\em cf.}
table \ref{tabelzm}). The known problem of negative effective mass
\cite{Levai,Prakash,Jurgen} manifests itself dramatically in panel
{\em a} for the Walecka model. We discuss this point in the next
subsection.

In figure \ref{fig4}  the populations in a system consisting of
hyperons, nucleons and leptons, for the Walecka and ZM models are
shown.  The poor description of neutron star phenomenology in the
ZM3 approach lead us to omit its results for the baryonic and
leptonic populations. Walecka's baryonic distribution stabilizes
after $\rho\sim 1.0fm^{-3}$ and all species appear up till
$\rho\sim0.7fm^{-3}$ which is approximately the density where
$|S|$ exceeds $M$. The lepton populations never vanish in the ZM
distribution and even at $\rho\sim1.2fm^{-3}$ baryonic species are
emerging. Essentially, these differences are due to the strength
of the scalar potential in these two models. Since a particle is
created only when
\begin{eqnarray}
q_B\mu_n\!-\! q_{e,B}\mu_e \!\ge\!  g_{\omega B}^\star \omega_0 \!
+\! g_{\rho
B}^\star \varrho_{03}I_{3B}\! +\!(M_B -g_{\sigma B}^\star\sigma) \,
\nonumber \\
\end{eqnarray}
a large scalar field favors the early emergence of  particles.
(One should compare theses results with the corresponding ones
shown in section \ref{3secpop} for the BB model.) In figure
\ref{4zmotv1} the predictions of the ZM model for the radial
distribution of the different lepton and baryon species is
presented.

We analyze now our new class of models allowing variations of the
$\lambda$ parameter. Tables  \ref{4tabste} and  \ref{4tabstev}
show results for the radius, {\em redshift} (z) and hyperon/baryon
ratio for the maximum mass of a neutron star sequence, considering
hyperons, nucleons and  leptons degrees of freedom, as a function
of $\lambda$; we present some nuclear matter properties as well.
In figure \ref{4mstlam} we show the dependence of the maximum mass
with this parameter. One can see that some models corresponding to
the S-V case (including ZM3 model discussed above) predict very
small neutron star masses, lower than the masses of all pulsars
found until now. We observe again the similarity of the
predictions associated to the ZM and ZM3 models compared to those
with exponential couplings, EXP/S and EXP/S-V.

The dependence of the maximum neutron star mass of a sequence with
the nuclear matter bulk properties $K$ and $M^\star$ at saturation
density is also presented in tables \ref{4tabste} and
\ref{4tabstev}.
Figure \ref{fig6} exhibits the dependence of the maximum mass of a
neutron star for a sequence with the compression modulus and
nucleon effective mass at saturation density in S and S-V cases.
 Figure \ref{fig3} shows the maximum neutron star mass
for a given value of the $\lambda$ parameter as a function of the
scalar potential $S$ in the center of the neutron star.

In both S and S-V cases, to a higher maximum
neutron star mass it corresponds, in general,  a less compressible
matter (higher $K$). To higher values of  the nucleon effective
mass it corresponds weaker scalar potentials; since the latter is
also directly related to the incompressibility of nuclear matter,
the maximum neutron star mass decreases with $M^\star$. In spite
that the two cases represent different descriptions of the neutron
star problem, the figure shows that to a fixed value of the
compression modulus it corresponds very close values for the
neutron star mass; opposite to that, to different values of the
star mass it correspond the same value for the nucleon effective
mass. This result indicates that the predictions of neutron star
masses based mainly on the  compression modulus are more model
independent than those based on the nucleon effective mass.

From figure \ref{fig3} and table \ref{4tabstev}, we see that for
$\lambda<0.5$ in the S-V case, we have obtained negative values
for the nucleon effective mass, which corresponds to a density
region for which the strength of the  scalar condensate exceeds
the free mass of the nucleon, $|S| > M$.

\subsection{Negative effective mass}
The nucleon effective mass is a dynamical quantity and expresses
the screening of the baryon masses by the scalar meson condensate.
By analyzing in our approach the general expression for the baryon
effective mass
\begin{eqnarray}
M^\star_B = M_B - \frac{g_\sigma \sigma}{(1+ g_\sigma \sigma /
(\lambda M))^\lambda} \label{eq:efma}
\end{eqnarray}
in the limit $g_\sigma\sigma\to \infty$, we see that only in case
$\lambda\ge1$ this quantity do not vanish or become negative. Some
internal constraints in the theory can avoid this, as is the well
known case of the Walecka model without hyperons \cite{walecka
86}. However, as we add more and more baryonic species we open the
possibility  for the scalar potential $|S|$ to become greater than
the free masses.
 Let us introduce the scalar field
equation in (\ref{eq:efma}), for the Walecka case, to get a better
understanding of the problem; we obtain
\begin{eqnarray}
M^\star_B \!\!= \!\!M_B\!\! - \!\!\sum_{B^\prime}
\!\frac{g_\sigma^2}{m_\sigma^2}\! \frac{M^\star_{B^\prime}}{\pi^2}
\!\! \int\!
\frac{k^2dk}{\sqrt{k^2\!+\!{M^\star_{B^\prime}}^2}}\,\,, \nonumber
\end{eqnarray}
or
\begin{eqnarray}
M^\star_B\!&=&\! \left(\!\!M_B\!\!-\!\!\sum_{B^\prime\ne B}\!
\frac{g_\sigma^2}{m_\sigma^2} \frac{M^\star_{B^\prime}}{\pi^2}
\!\! \int\!
\frac{k^2dk}{\sqrt{k^2\!+\!{M^\star_{B^\prime}}^2}}\!\!\right)\!/
\nonumber\\ && \left(1+ \left(\frac{g_\sigma}{m_\sigma}\right)^2
\frac{1}{\pi^2}  \int
\frac{k^2dk}{\sqrt{k^2+{M^\star_{B}}^2}}\right)\,\,.
\label{effect}
\end{eqnarray}
As we add more and more baryonic species, the negative term in the
numerator of (\ref{effect}) becomes more important and we open the
possibility for the scalar potential $|S|$ to become greater than
the free baryon masses. In fact, even if we had just considered
nucleons, but taking into account the difference in the neutron
and proton masses, that negative term in the numerator would
appear. However, since this difference is very tiny, a negative
effective mass would emerge  in practice  only with the addition
of hyperon degrees of freedom \cite{Jurgen}.

We could interpret the vanishing  of the effective mass as a
signal of a transition to a quark-gluon plasma phase. However, we
should remember that our Lagrangian model does not comprehend
these underlying degrees of freedom. Additionally, as pointed out
in \cite{mishustin}, at such high densities and strong meson
fields we have already reached the critical  density where the
production of baryon-antibaryon pairs is favored. In fact, this
behavior of the effective mass may indicate that the mean field
approximation is reaching  the limits of its applicability.

Concerning the problem of negative effective mass, we want to
clarify a point which may appear naive but has led to
misunderstandings. The integrals appearing in the RMF formalism
always present a term ${M^\star}^2$ and, in this way, are
symmetric with respect to positive or negative values of the
effective mass. For example, let us take the integral related to
the scalar density:
\begin{eqnarray}
I(t,m)=\int_0^t\frac{k^2dk}{\sqrt{k^2+m^2}};
\end{eqnarray}
rigorously, the expression for $I(t,m)$ would involve the modulus
of $m$ since $\sqrt{m^2}=|m|$:
\begin{eqnarray} \int_0^t\!\!\frac{k^2}{\sqrt{k^2+m^2}} \!\!&=&\!\! \frac12
t\sqrt {{t}^{2}\!+\!{{{m}}}^{2}}\!\!-\!\!\frac12{ m}^{2}\ln
\left(\frac{t+\sqrt{{t}^{2}\!+\!{m}^{2}}}{|m|}\right).\nonumber
\\&& \label{negative}
\end{eqnarray}
Some authors have assumed {\em ad hoc} the modulus of $M^\star$ in
the above logarithm  when the problem of negative values appears.
However, we can see from expression (\ref{negative}) that it
emerges naturally from the symmetry of the integrals. Of course,
different results would arise if we are using $|M^\star|$ instead
of $M^\star$ in other expressions, {\em e.g.} the cubic term in
the energy density expression for the BB model, since it can be
rewritten as
\begin{eqnarray}
\frac13bM(M-M^\star)^3.
\end{eqnarray}
Finally, we want to stress out that, in spite of the interesting
issue about the  physical interpretation of a negative effective
mass, mathematically the RMF  model continues to work even with
$M^\star<0$.

\section{Summary and Conclusions}
We have done an analysis on the influence of nuclear matter
properties on the structure of neutron stars using a new class of
RMF models with parameterized couplings among mesons and baryons.
These couplings allow to reproduce, trough a suitable choice of
mathematical parameters ($\lambda,\beta,\gamma$), results of the
Walecka and derivative coupling models such as ZM and ZM3 models.
As we have shown above, this new class of relativistic models
permits some control on the intensity of the scalar, vector and,
isovector mesons mean-field potentials. In particular, we have
scanned
 values of nuclear matter properties which correspond to intermediate
region of predictions of the Walecka, ZM  and ZM3 models.

In this new class of models we considered two cases, the Scalar
(case S) with variations of $\lambda$ and $\beta = \gamma = 0$ (
this case contains the results of the Walecka and  ZM models) and
the Scalar-Vector (case S-V) with variations of $\lambda$ and
$\beta = \gamma  = \lambda$ (ZM3 model belongs to this category).
Taking the limit $\lambda\rightarrow\infty$ for the S case, we
have a model with exponential couplings (EXP/S model). Similarly,
for $\lambda = \beta = \gamma \rightarrow\infty$ we have the
asymptotic model EXP/S-V.

An extended RMF which includes hyperon and lepton degrees of
freedom was presented. For the sake of simplicity, we have assumed
an universal coupling for the meson-baryon couplings. We have
investigated the effects of the non-linear coupling in the nucleon
effective mass and compression modulus of nuclear matter and in
neutron star properties.

Having this in mind let us summarize the main findings of our
work:

$\bullet$ For the nuclear matter properties we have found values
of parameters that reproduce reasonably the best range of
phenomenological results, namely $0.6 < M^*/M < 0.7$ and $200 < K
< 300 $ MeV. We have also found  that to higher values of the
nucleon effective mass it corresponds lower values of the
compression modulus. The Scalar case leads to a narrow interval of
values of $\lambda$ with different descriptions of nuclear matter
properties  while the Scalar-Vector case has a broader range. To
be more specific, case S gives reasonable results in the range
$\lambda \sim 0.05 - 0.07$ and case S-V better results for
$\lambda \sim 0.16 - 0.4$. We have also pointed out the strong
similarity between the results of ZM-like models and those with
exponential couplings. We have also obtained  an analytical
expression
 for the compression modulus of nuclear matter as a function
of the nucleon effective mass (Appendix \ref{appendixb}).

$\bullet$ We have pre\-sen\-ted ex\-pli\-ci\-tly the neu\-tron
star phe\-no\-me\-no\-lo\-gy for the Wa\-le\-cka, ZM and ZM3
models. To the best of our knowledge, we display in this work, for
the first time, the results of these two later models. The ZM
model predicts a maximum mass of approximately $1.6 M_\odot$ for a
neutron star, while the ZM3, being too soft, leads to  $\sim0.72
M_\odot$ as the limiting mass. We have obtained Walecka and ZM
fermionic distributions. The corresponding particle populations
exhibit expressive differences essentially due to the strength of
the scalar potential in these two models.

$\bullet$ Sensibility of this approach to the specific choice of
$B$ and $\rho_0$ was noticed with differences of the order of a
half solar mass.  For the Walecka model, the neutron star mass
with hyperons has varied from $2.77M_\odot$ to $2.33M_\odot$ with
a very small change in those nuclear matter bulk properties. Thus,
extrapolation for neutron star densities from the fitting of $B$
and $\rho_0$ at saturation needs more precision on the choice of
the values for these quantities.

$\bullet$  We analyzed our new class of models allowing the variation of
 the $\lambda$ parameter and  showed results for the
radius, redshift (z) and hyperon/baryon ratio for the maximum mass
of a neutron star sequence, considering hyperon, nucleon and
lepton degrees of freedom. In figure \ref{4mstlam} we showed the
dependence of the maximum mass with this parameter. We found that
some models corresponding to the S-V case (including the ZM3
model) predicted very small neutron star masses. We observed again
the similarity between  the predictions of neutron star
phenomenology associated to the ZM and ZM3 models compared to
those with exponential couplings, EXP/S and EXP/S-V.

$\bullet$ The dependence of the maximum neutron star mass of a sequence with
the nuclear matter bulk properties $K$ and $M^\star$ at saturation
density was also analyzed.
In both S and S-V cases, to a higher maximum
neutron star mass it corresponds, in general,  a less compressible
matter (higher $K$). We also concluded that the maximum neutron star mass decreases with
the nucleon effective mass $M^\star$.

$\bullet$ Finally, we found that for $\lambda<0.5$ in the S-V
case, we have obtained negative values for the nucleon effective
mass, which corresponds to a density region for which the strength
of the  scalar condensate exceeds the free mass of the nucleon, $|
S| > M$. We have shown how the differences in the baryonic bare
masses explain this result; as explained in the text, in practice
the inclusion of hyperons is responsible for this behaviour. These
results indicate the existence of very intense scalar condensates
in the interior region of neutron stars. Concerning  mathematics,
the theory continues to work due to the symmetry of negative and
positive values of $M^\star$ in the Fermi integrals.

\acknowledgments This work has been supported by CNPq.

\appendix

\section{ Chemical equilibrium} \label{appendixa}

Chemical reactions,{\it e.g.} $ A+B \rightleftharpoons C + D $,
can be expressed
 in general,
as a symbolic linear combination
of its components \cite{landau}:
\begin{eqnarray}
\sum_{i}\nu_i A_i=0.
\label{equi1}
\end{eqnarray}
For example, in the reaction $n \rightleftharpoons p^+ + e^- +
\bar{\nu}$ we have for the coefficients $\nu_{n}= -
\nu_{p}=-\nu_{e}=-\nu_{\nu} = 1$.

We shall consider infinitesimal variations of the Gibbs potential
($ G \equiv G(p,T,N_i)$), with respect to the number of particles
($N_i$), at constant temperature T and pressure p
\begin{eqnarray}
dG = \sum_i{\left(\frac{\partial G}{\partial N_i}\right)}_{_{T,p,
N_{j \ne i}}} \!\!\!\!\! dN_i \, .
\end{eqnarray}
At chemical equilibrium, Gibbs energy obeys the condition:
\begin{eqnarray} \sum_j{\left(\frac{\partial G}{\partial N_j}\right)}_{_{T,p,
N_{j \ne i}}} \!\!\!\!\! \frac{dN_j}{dN_i} = 0 \, ,
\end{eqnarray}
with the ratio $\frac{dN_j}{dN_i}$ determined by the corresponding
chemical reactions. Accordingly, if an element $i$ suffers a
variation $\bar{\nu}_i$, the remaining elements will suffer a
variation $(\bar{\nu_i}/\nu_i) \nu_j$ to maintain the
stoichiometry of the reaction. As a result, we have $dN_j/dN_i =
\nu_j / \nu_i$ and we can write the condition of chemical
equilibrium: \begin{eqnarray} \sum_i\nu_i \mu_i = 0, \label{equi2}
\end{eqnarray}
where the chemical potential of element $i$, $\mu_i$, is defined
as: \begin{eqnarray} \mu_i \equiv {\left(\frac{\partial
G}{\partial N_i}\right)}_{_{T,p, N_{i \ne j}}}.
\end{eqnarray}In this way we observe
that the chemical potentials obey the symbolic equation
(\ref{equi1}), with the substitution of $A_i$ by $\mu_i$.

In general, if a chemical reaction respects given conservation
laws,  the number of independent chemical potentials is equal to
the number of these laws.

In the following we consider two conservation laws, electric
charge and baryon number. In this case we can express these laws,
for a chemical reaction,
as: \begin{eqnarray} \sum_i^N \nu_i q_{ei} = 0 \,\,\,\,\,\,\,\,\,
\textrm{and} \,\,\,\,\,\,\, \sum_i^N \nu_i q_{bi}=0,
\end{eqnarray}where $q_{ei}$ and  $q_{bi}$ denote,
respectively, the electric and baryon  charges of element $i$. As
we have, in this case,
 $N$ variables and  two equations, we are able
to express only two coefficients $\nu_i$ in terms of the remaining
$N-2$, which will be independent: \begin{eqnarray} \nu_1 q_{e1} +
\nu_2 q_{e2} = -\sum_{i\ne 1,2}^N \nu_i q_{ei}, \\ \nu_1 q_{b1} +
\nu_2 q_{b2} = -\sum_{i \ne 1,2}^N \nu_i q_{bi}.
\end{eqnarray}
We consider as an example element $1$, the  neutron,  and element
$2$, the electron.  We then have $q_{b1} = -q_{e2} = 1$, $q_{b2} =
q_{e1}=0$ and the above equations become \begin{eqnarray} \nu_n =
-\sum_{i \ne n,e} ^N \nu_{i} q_{bi}, \nonumber \\ \nu_e = \sum_{i
\ne n,e}^N \nu_{i} q_{ei}. \label{equi3}
\end{eqnarray}Replacing (\ref{equi3}) in (\ref{equi2}) we find:
\begin{eqnarray}
\sum_{i \ne n,e}^N \nu_i \mu_i = \sum_{i \ne n,e}^N (\mu_n q_{bi}) \nu_i
- \sum_{i \ne n,e}^N (\mu_e q_{ei}) \nu_i.
\end{eqnarray}
Since the $\nu_i$ are  independent, the equality of this equation
will be verified only if the coefficients are equal. From this
expression then results
 \begin{eqnarray} \mu_i = q_{bi}\mu_n - q_{ei}\mu_e.
\label{equi4}
\end{eqnarray}

\section{Compression modulus} \label{appendixb}

 From the definition of the compression modulus,
\begin{eqnarray} K =\left[ k^2 \frac{d^2 (\varepsilon/\rho)}{dk^2}\right]_{\varrho
=\varrho_0}  ,
\end{eqnarray}
we have \begin{eqnarray} K&=& \left[k^2\left(
\frac{p}{\rho}\frac{d^2 \rho}{dk^2} +
{\left(\frac{d\rho}{dk}\right)}^2 \left(\frac{-2p}{\rho^3}
+\frac{1}{\rho}
\frac{d\mu}{d\rho}\right)\right)\right]_{\rho=\rho_0}. \nonumber \\
&&\label{K}
\end{eqnarray}
At saturation density, the pressure of nuclear matter
is null. From expression (\ref{K}) we
then have
\begin{eqnarray}
K = \left. 3k_0{\frac{d\mu}{dk}} \right|_{k=k_0}  .
\label{k}
\end{eqnarray}
The chemical potential is a function
of $k$ and
$\sigma$. Thus, its derivative
with respect to  $k$ is:
\begin{eqnarray}
\frac{d\mu}{dk} =  \frac{\partial \mu}{\partial \sigma} \frac{d\sigma}{dk}
+ \frac{\partial \mu}{\partial k}.
\label{kmu}
\end{eqnarray}
From equation (\ref{potlamb}) for the chemical potential we have
for the last term of expression (\ref{kmu}) \begin{eqnarray}
\frac{\partial \mu}{\partial k} = {m^*_\beta}^2\frac{g_\omega^2}
{m_\omega^2} \frac{2k^2}{\pi^2}+\frac{k}{\sqrt{k^2+{M^*}^2}} \, ,
\label{kk}
\end{eqnarray}
and for the derivative of the chemical potential
with respect to the $\sigma$ field we get:
\begin{eqnarray}
\frac{\partial \mu}{\partial \sigma} = 2 m^*_\beta
\frac{g_\omega^2}{m_\omega^2}\frac{\partial m^*_\beta}{\partial \sigma}
\frac{2k^3}{3 \pi^2}+\frac{M^*}{\sqrt{k^2+{M^*}^2}}
\frac{\partial M^*}{\partial \sigma}.
\label{ksig}
\end{eqnarray}
To find the contribution of the $d\sigma/dk$ term in expression
(\ref{kmu}), we consider a profile function $f=f(\sigma,k)$, and
the condition \begin{eqnarray} f(\sigma,k) \equiv m_\sigma^2
\sigma + F(\sigma) \rho_s+G(\sigma) m^*_\beta
\frac{g_\omega^2}{m_\omega^2} \rho^2 = 0 \, ;
\end{eqnarray}
from this expression
\begin{equation}
\frac{d\sigma}{dk} = \frac{-(\partial f/\partial k)_\sigma}{(\partial f/
\partial \sigma)_k} \, ,
\end{equation}
with
\begin{equation}
{\left(\frac{\partial f}{\partial k}\right)}_\sigma =
F(\sigma) \frac{\partial \rho_s}{\partial k} + 2 G(\sigma) m^*_\beta
\frac{g_\omega}{m^2_\omega} \rho \frac{d\rho}{dk} \, ,
\label{kf1}
\end{equation}
and \begin{eqnarray} {\left(\frac{\partial f}{\partial
\sigma}\right)}_k & \! = \! &  m^2_\sigma + \frac{\partial
F(\sigma)} {\partial \sigma} \rho_s + F(\sigma) \frac{\partial
\rho_s}{\partial \sigma} + m^*_\beta \frac{g_\omega}{m_\omega^2}
\frac{\partial G(\sigma)} {\partial \sigma} \rho^2 \nonumber \\ &
\! + \! & \frac{G(\sigma)}{m^2_\omega} \frac{\partial
g^\star_\omega}{\partial \sigma} \rho^2 \, . \label{kf2}
\end{eqnarray}
According to equations (\ref{fg1})
and (\ref{fg2}),
the derivatives of the
$F(\sigma)$ and $G(\sigma)$
functions
are
\begin{eqnarray}
\frac{\partial F(\sigma)}{\partial \sigma} =
g_\sigma^2\left[ \frac{2}{M}{m^*_\lambda}^{\frac{\lambda+1}{\lambda}}
-\frac{\lambda+1}{\lambda}\frac{g_\sigma\sigma}{M^2}
{m^*_\lambda}^{\frac{\lambda+2}{\lambda}} \right] \, , \nonumber \\
\label{dfsig}
\end{eqnarray}
and
\begin{eqnarray}
\frac{\partial G(\sigma)}{\partial \sigma} = g_\omega g_\sigma^2
\left[\frac{\beta+1}{\beta} \frac{1}{M^2}
{m^*_\beta}^{\frac{\beta+2}{\beta}}\right] \, .
\label{dgsig}
\end{eqnarray}
To obtain an equation for the compression modulus of nuclear
matter as a function of $g_\sigma/m_\sigma$ and
$g_\omega/m_\omega$, we have to consider the following
ratios \begin{eqnarray} \frac{F(\sigma)}{g_\sigma}, \,\,\,
\frac{G(\sigma)}{g_\sigma g_\omega}, \,\,\,
\frac{1}{g_\sigma^2}\frac{\partial F(\sigma)}{\partial \sigma}
\,\,\, \textrm{and} \,\,\, \frac{1}{g_\omega g_\sigma^2}
\frac{\partial G(\sigma)}{\partial \sigma} \, .
\end{eqnarray}
Combining
(\ref{kk}), (\ref{ksig}), (\ref{kf1})
and (\ref{kf2}) with (\ref{kmu}), we
finally obtain:
\begin{eqnarray}
K = K_{_{1}} + K_{_{2}} + \frac{K_{_{3A}} \times K_{_{3B}}}{K_{_{3C}}},
\end{eqnarray}
where
\begin{eqnarray}
K_{_{1}} = {m^*_\beta}^2 \left(\frac{g_\omega^2}{m_\omega^2}\right) \frac{6 k^3_0}{\pi^2}, \\
\nonumber \\
K_{_{2}} = \frac{3k_0^2}{\sqrt{k_0^2+{M^*}^2}} \, ,
\end{eqnarray}
\begin{eqnarray}
K_{_{3A}} & = &  -3 K_0\left[2\left(\frac{g_\omega^2}{m_\omega^2}\right)m^*_\beta
\left(\frac{G(\sigma)}{g_\sigma g_\omega}\right) \frac{2 k_0^3}{3 \pi^2}
\right. \nonumber \\
 & + & \left. \frac{M^*}{\sqrt{k_0^2+{M^*}^2}}\left(\frac{F(\sigma)}{g_\sigma}
\right)\right],
\end{eqnarray}
\begin{eqnarray}
K_{_{3B}} &  =    & \left[ \! \left(\frac{g_\sigma^2}{m_\sigma^2}\right)
\left(\frac{F(\sigma)}{g_\sigma}\right)  \frac{2 k^2_0 M^*}
{\sqrt{k_0^2+{M^*}^2}}\right] \nonumber \\
&  \times   &
\left[1  +   \frac{\left[2 m^*_\beta  \left(\frac{G(\sigma}{g_\sigma g_\omega}\right)
\left(\frac{g_\omega^2}{m^2_\omega}\right)
 \frac{4k_0^5}{3 \pi^4}\right]}
{\left[\frac{F(\sigma)}{g_\sigma}\frac{2k^2_0M^*}{\sqrt{k_0^2+{M^*}^2}}
\right]} \right] \, ,
\end{eqnarray}
and
\begin{eqnarray}
K_{_{3C}} &  \! = \!   &  1 \!  + \!
\left(\frac{g_\sigma^2}{m_\sigma^2}\right)
\left(\frac{1}{g_\sigma^2}\frac{\partial F(\sigma)}{\partial \sigma}\right)
\! \left( \!  \frac{2}{\pi^2}\int_0^{k_0} \!  \frac{M^* k^2}
{\sqrt{k_0^2  \! + \!  {M^*}^2}}dk \! \right) \nonumber \\
& \! +  \! &
 {\left(\frac{F(\sigma)}{g_\sigma}\right)}^2
 \left(\frac{g_\sigma^2}{m_\sigma^2}
\right)  \left(\frac{2}{\pi^2}\int_0^{k_0}  \frac{k^4_0}{(k_0^2
 \! + \!
{M^*}^2)^{3/2}}dk\right) \nonumber \\
&  \! + \!   & \left(\frac{g_\omega^2}{m_\omega^2}\right)
 \left(\frac{g_\sigma^2}{m_\sigma^2}\right)
 \frac{4 k_0^6}{9 \pi^4}
\nonumber \\
& \! \times \! &
 \left[ m^*_\beta \left(  \frac{1}{g_\omega
g_\sigma^2} \frac{\partial G(\sigma)}{\partial \sigma}\right)
 \! + \!   {\left(\frac{G(\sigma)}{g_\sigma g_\omega}\right)}^2
 \right] \, . \nonumber \\
\end{eqnarray}


\newpage

\mediumtext
\begin{table}
\caption{Values of coupling constants, nucleon effective mass,
compression modulus of nuclear matter and scalar and vector potentials
at saturation density (ZM and ZM3 models). \label{tabelzm}}
\begin{tabular}{l c c c c c c c}
Model  & $(g_\sigma/m_\sigma)^2$ & $(g_\omega/m_\omega)^2$ &
$(g_\rho/m_\rho)^2$&$M^\star/M$ & $K $ & S & V   \\
&$(fm^2)$&$(fm^2)$&$(fm^2)$&&(MeV)&(MeV)&(MeV) \\ \tableline ZM &
7.94 & 2.84 & 5.23 & 0.85 & 224 & -140 & 84 \\ ZM3 & 19.57 & 13.45
& 9.06 & 0.71 & 159 & -267& 204  \\
\end{tabular}
\end{table}

\narrowtext
\begin{table}
\caption{Nucleon effective mass and compression modulus of nuclear
matter for different type of couplings between the scalar mesons
and the nucleon fields {\protect \cite{koepf}}.\label{tabelfe}}
\begin{tabular}{l c c c}
&$\bar{m}^\star(\sigma)$  & $M^*/M$ & $K (MeV)$ \\ \tableline
1 &
$1-\frac{g_\sigma\sigma}{M}$ &  0.55 & 545 \\ 2 &
$1-tanh(\frac{g_\sigma}{M}\sigma)$ & 0.71 & 410 \\ 3 &
$exp(-\frac{g_\sigma \sigma}{M})$ & 0.80 & 265 \\ 4 &
$(1+\frac{g_\sigma \sigma}{M})^{-1}$ & 0.85 & 233 \\
\end{tabular}
\end{table}

\begin{table}
\caption{Values of $\lambda$, $\beta$ and $\gamma$ for different
QHD models. \label{corres}}
\begin{tabular}{l c c c}
Model&$\lambda$  & $\beta$ & $\gamma$  \\ \tableline Walecka & $0$ &
$0$ & $0$ \\ ZM & $1$ & $0$ & $0$  \\ ZM3 & $1$ & $1$ & $1$ \\
\end{tabular}
\end{table}

\protect\newpage
\mbox{}
\protect\newpage

\widetext
\begin{table}
\caption{Stellar properties for the S case: $\varepsilon_c$ -
central density;$M_\star$ star mass; $R_\star$ - star radius; $S$ - scalar
potential in the star center; $z$ - redshift; $Y/A$ is the
hyperon/baryon ratio and  $N_{BT}$ is the total baryonic number.
All this quantities are evaluated  for the neutron star with the
maximum mass in the sequence.Besides we have the compression
modulus $K$, the nucleon effective mass $M^\star/M$ and the
relativistic coefficient $R$. \label{4tabste}}
\begin{tabular}{ c c c c c c c c c c c }
$\lambda$&$log(\varepsilon_c)$&$M_\star$& $R_\star$ & S    &  z&
Y/A & $N_{BT}$    & K& $M^\star/M$&R\\
&$g/cm^3$&($M_\odot$)&(km)&(MeV)&&&($\times 10^{58}$)&(MeV)&&\\
\tableline
 0&      15.18&       2.77&      13.17&      936&      0.623&       0.27&       0.40&     566&      0.537&      0.931\\
  0.03&      15.24&       2.56&      12.39&      934&      0.597&       0.30&       0.36&     396&      0.598&      0.943\\
  0.05&      15.31&       2.35&      11.63&      929&      0.574&       0.32&       0.33&     310&      0.650&      0.951\\
  0.07&      15.38&       2.17&      10.89&      923&      0.554&       0.34&       0.30&     258&      0.694&      0.957\\
  0.09&      15.43&       2.02&      10.38&      910&      0.533&       0.35&       0.28&     235&      0.725&      0.960\\
 0.11&      15.47&       1.91&       9.98&      896&      0.515&       0.36&       0.26&     223&      0.749&      0.962\\
 0.13&      15.49&       1.83&       9.75&      877&      0.495&       0.35&       0.25&     217&      0.766&      0.964\\
 0.15&      15.52&       1.77&       9.59&      857&      0.479&       0.35&       0.24&     216&      0.779&      0.965\\
 0.17&      15.54&       1.72&       9.45&      838&      0.467&       0.35&       0.23&     213&      0.789&      0.966\\
 0.20&      15.53&       1.68&       9.48&      807&      0.446&       0.33&       0.23&     212&      0.798&      0.967\\
 0.25&      15.53&       1.61&       9.49&      732&      0.416&       0.30&       0.22&     212&      0.814&      0.968\\
 0.30&      15.52&       1.59&       9.61&      669&      0.399&       0.27&       0.21&     214&      0.822&      0.969\\
 0.35&      15.51&       1.58&       9.69&      618&      0.389&       0.26&       0.21&     216&      0.828&      0.969\\
 0.40&      15.51&       1.58&       9.73&      580&      0.385&       0.25&       0.21&     218&      0.833&      0.969\\
 0.60&      15.49&       1.58&       9.86&      480&      0.377&       0.22&       0.21&     223&      0.843&      0.970\\
1.00&      15.47&       1.59&       9.98&       401&      0.372&       0.20&       0.21&     224&      0.850&      0.970 \\
1.50&      15.47&       1.59&       9.98&       366&      0.373&       0.20&       0.21&     226&      0.854&      0.971 \\
$\infty$& 15.47&       1.59&      10.00&        350&      0.373&       0.20&       0.21&     228&      0.856&       0.971\\
\end{tabular}
\end{table}

\widetext
\begin{table}
\caption{Stellar properties for the S-V case. Same correspondences
as in table \ref{4tabste}. \label{4tabstev}}
\begin{tabular}{ c c c c c c c c c c c}
$\lambda$&$log(\varepsilon_c)$&$M_\star$& $R_\star$ & S    &  z&
Y/A & $N_{BT}$    & K& $M^\star/M$&R\\
&$g/cm^3$&($M_\odot$)&(km)&(MeV)&&&($\times 10^{58}$)&(MeV)&&\\
\tableline
  0&      15.18&       2.77&      13.17&      936&      0.623&       0.27&       0.40&     566&      0.537&      0.931\\
  0.03&      15.20&       2.70&      12.93&      944&      0.615&       0.28&       0.39&     510&      0.545&      0.933\\
  0.05&      15.22&       2.63&      12.64&      954&      0.610&       0.29&       0.38&     458&      0.554&      0.935\\
  0.07&      15.24&       2.56&      12.39&      960&      0.602&       0.30&       0.37&     417&      0.561&      0.936\\
  0.09&      15.27&       2.50&      12.12&      969&      0.598&       0.31&       0.36&     387&      0.567&      0.938\\
 0.11&      15.28&       2.43&      11.89&      973&      0.588&       0.32&       0.34&     358&      0.574&      0.939\\
 0.13&      15.30&       2.37&      11.68&      977&      0.579&       0.33&       0.33&     339&      0.579&      0.940\\
 0.15&      15.33&       2.30&      11.38&      985&      0.574&       0.34&       0.32&     311&      0.587&      0.941\\
 0.17&      15.35&       2.24&      11.16&      986&      0.563&       0.34&       0.31&     293&      0.594&      0.942\\
 0.20&      15.38&       2.17&      10.88&      993&      0.559&       0.35&       0.30&     276&      0.600&      0.943\\
 0.30&      15.51&       1.83&       9.58&     1011&      0.516&       0.39&       0.25&     218&      0.630&      0.949\\
 0.35&      15.56&       1.70&       9.09&     1012&      0.491&       0.41&       0.23&     205&      0.640&      0.950\\
 0.40&      15.62&       1.57&       8.60&     1014&      0.470&       0.42&       0.21&     195&      0.649&      0.951\\
 0.60&      15.62&       1.07&       8.08&      891&      0.282&       0.35&       0.14&     169&      0.682&      0.955\\
1.00&      15.31&       0.72&       9.76&      577&      0.128&        0.10&       0.09&     159&      0.710&      0.959\\
1.50&      15.18&       0.67&      10.21&      468&      0.113&        0.04&       0.08&     156&      0.728& 0.961\\
$\infty$&      15.14&       0.66&      10.31&  431& 0.110&         0.03&       0.08&     155&      0.738& 0.961\\
\end{tabular}
\end{table}

\protect\newpage
\mbox{}
\protect\newpage

\onecolumn

\newpage

\newpage

\newpage

\newpage

\newpage

\begin{figure}
\caption{Dependence of the coupling
constants with the  $\lambda$ parameter for the scalar (full
lines) and scalar-vector (dotted lines) models. \label{figcon1}}
\centerline{\epsfysize=8cm \epsfbox{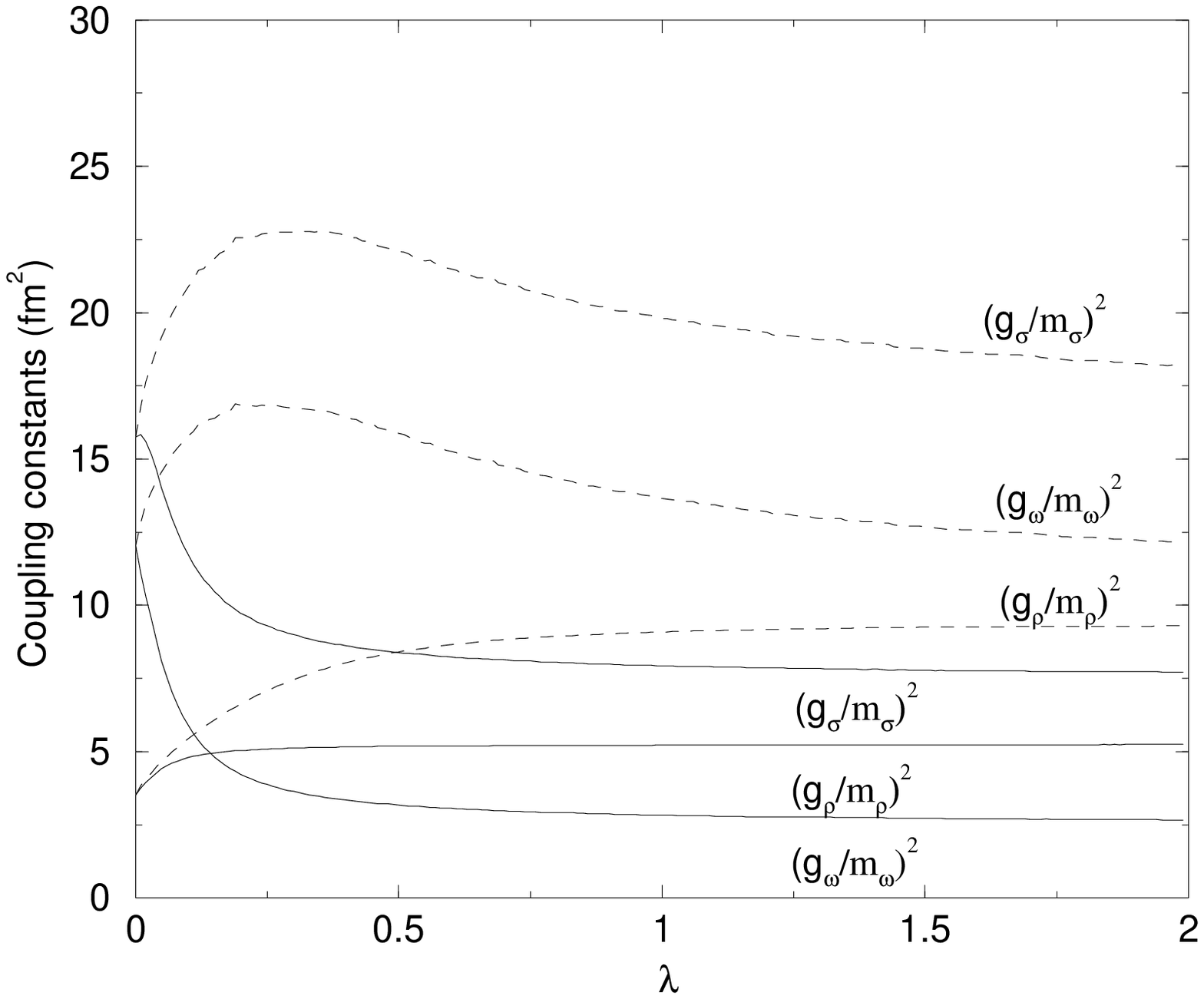}}
\end{figure}

\begin{figure}
\caption{Plane  of the scalar $S$ and vector
$V$ potentials. The potentials in the
Scalar and Scalar-Vector exponential coupling models are indicated." \label{figcon2}}
\centerline{\epsfysize=8cm \epsfbox{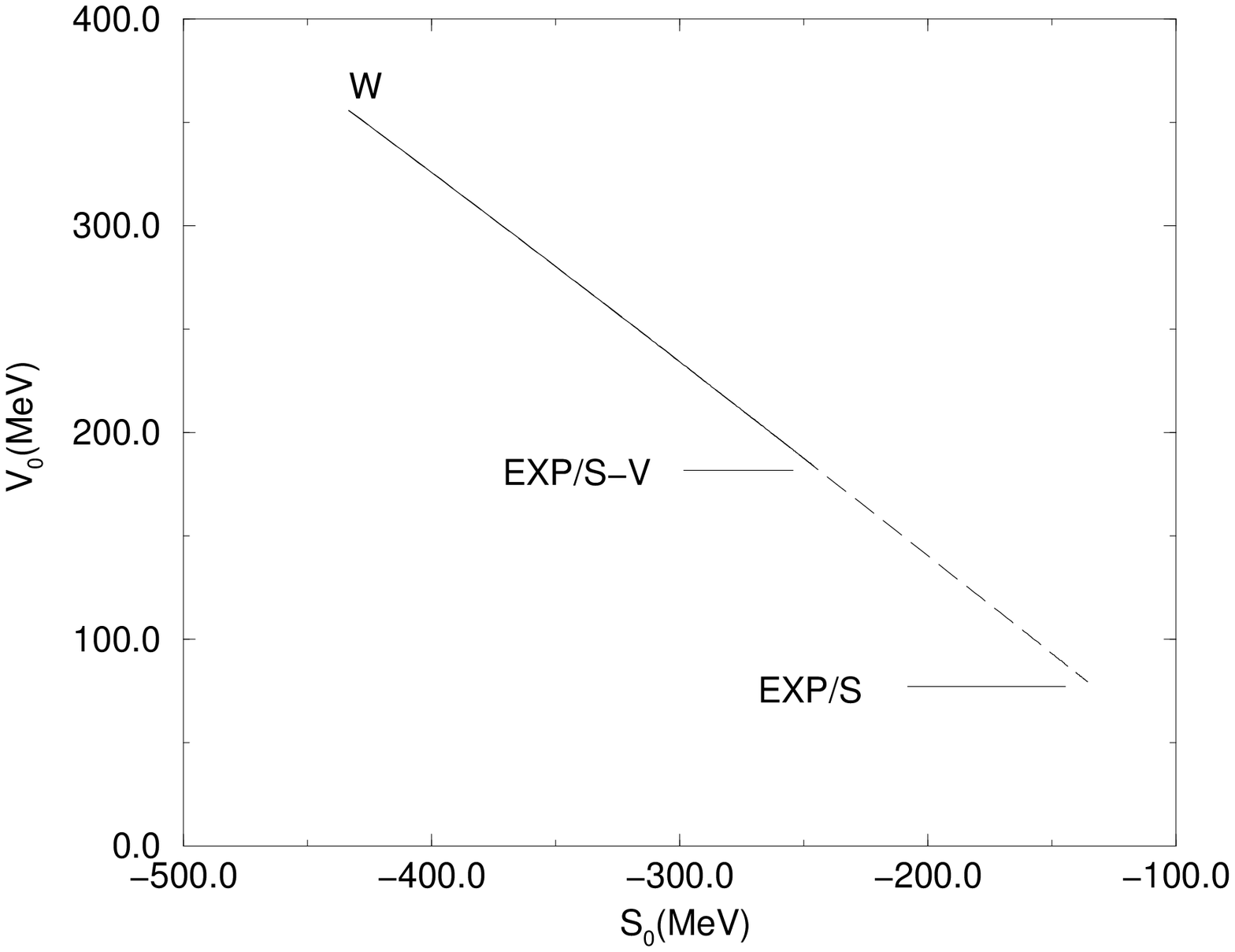}}
\end{figure}

\newpage
\begin{figure}
\caption{Compression modulus of nuclear matter
$K$ as a function of the $\lambda$ parameter.} \label{4kmlam1}
\centerline{\epsfysize=8cm\epsfbox{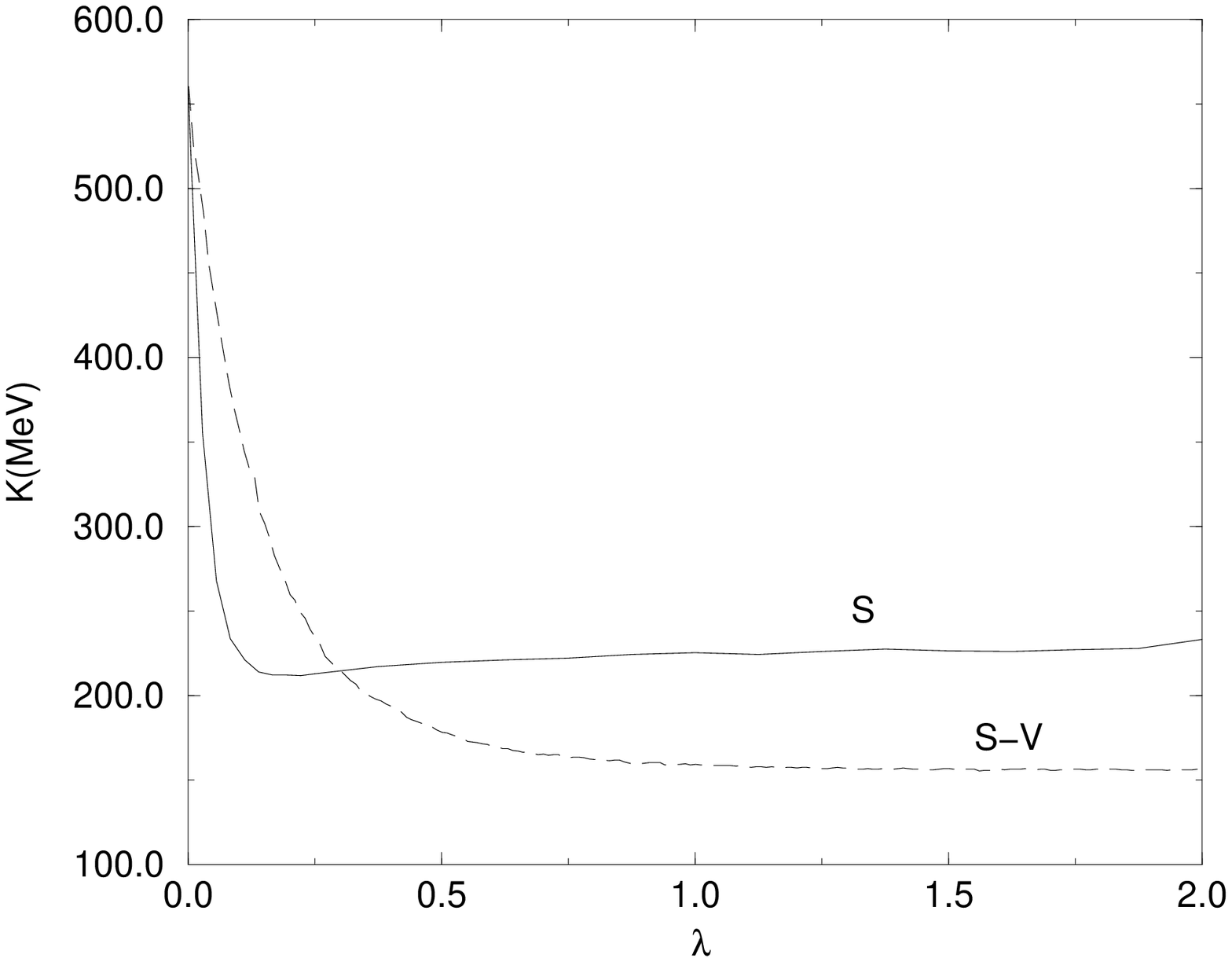}}
\end{figure}

\begin{figure}
\caption{The ratio $M^*/M$, at saturation
density, as a  function of the $\lambda$ parameter.}
\label{4kmlam2}
\centerline{\epsfysize=8cm \epsfbox{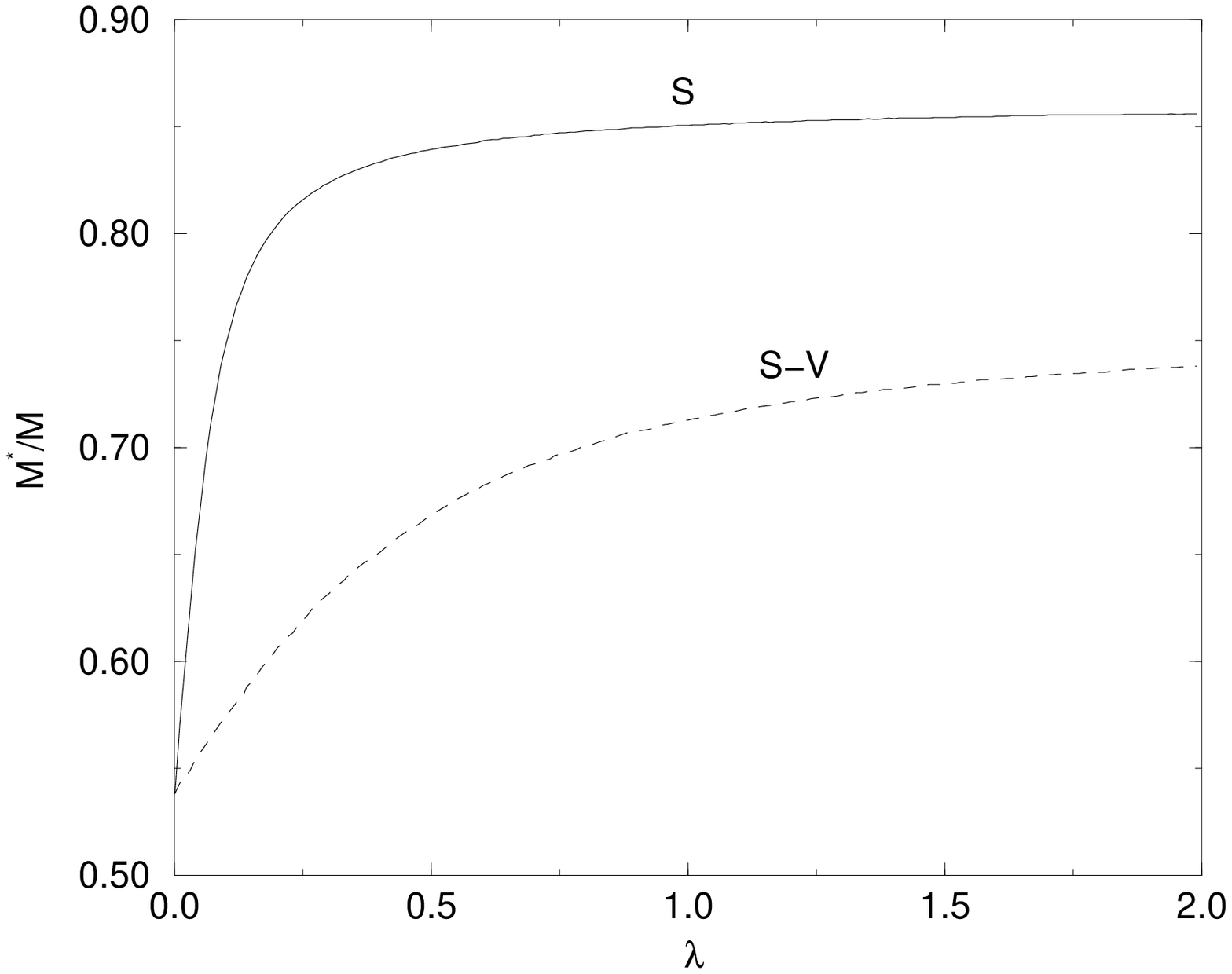}}
\end{figure}

\newpage

\begin{figure}
\caption{Comparison of the compression
modulus of nuclear matter $K$ with the ratio $M^*/M$ for the S and
S-V cases (the values for the exponential models are indicated). The box shows the range of accepted values.}
\label{4km1}
\centerline{\epsfysize=8cm \epsfbox{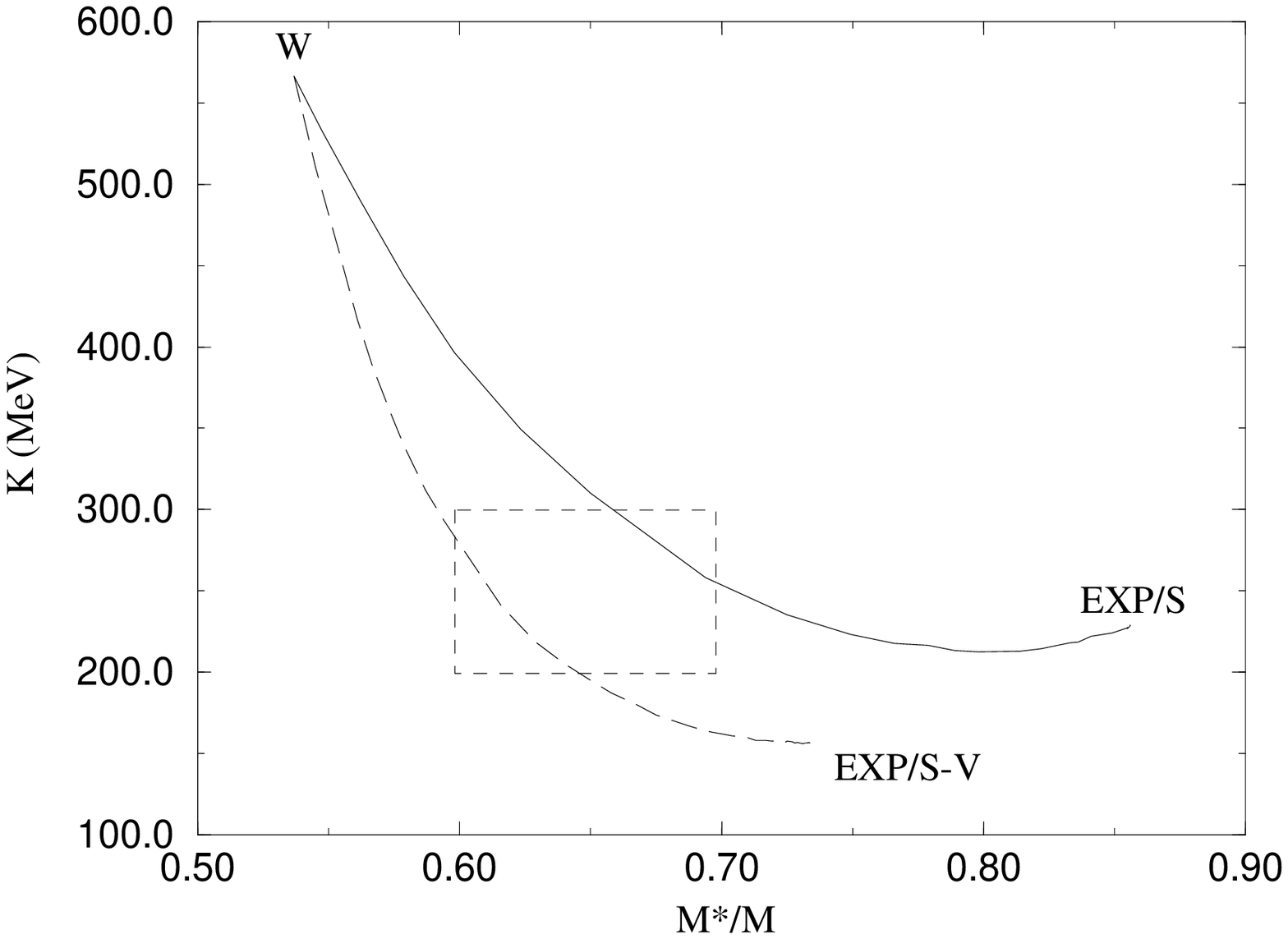}}
\end{figure}

\newpage

\begin{figure}
\caption{The ratio $M^*/M$ as a function of
the relativistic coefficient  $R$ for the S and S-V cases.
\label{4km2}}
\centerline{\epsfysize=8cm \epsfbox{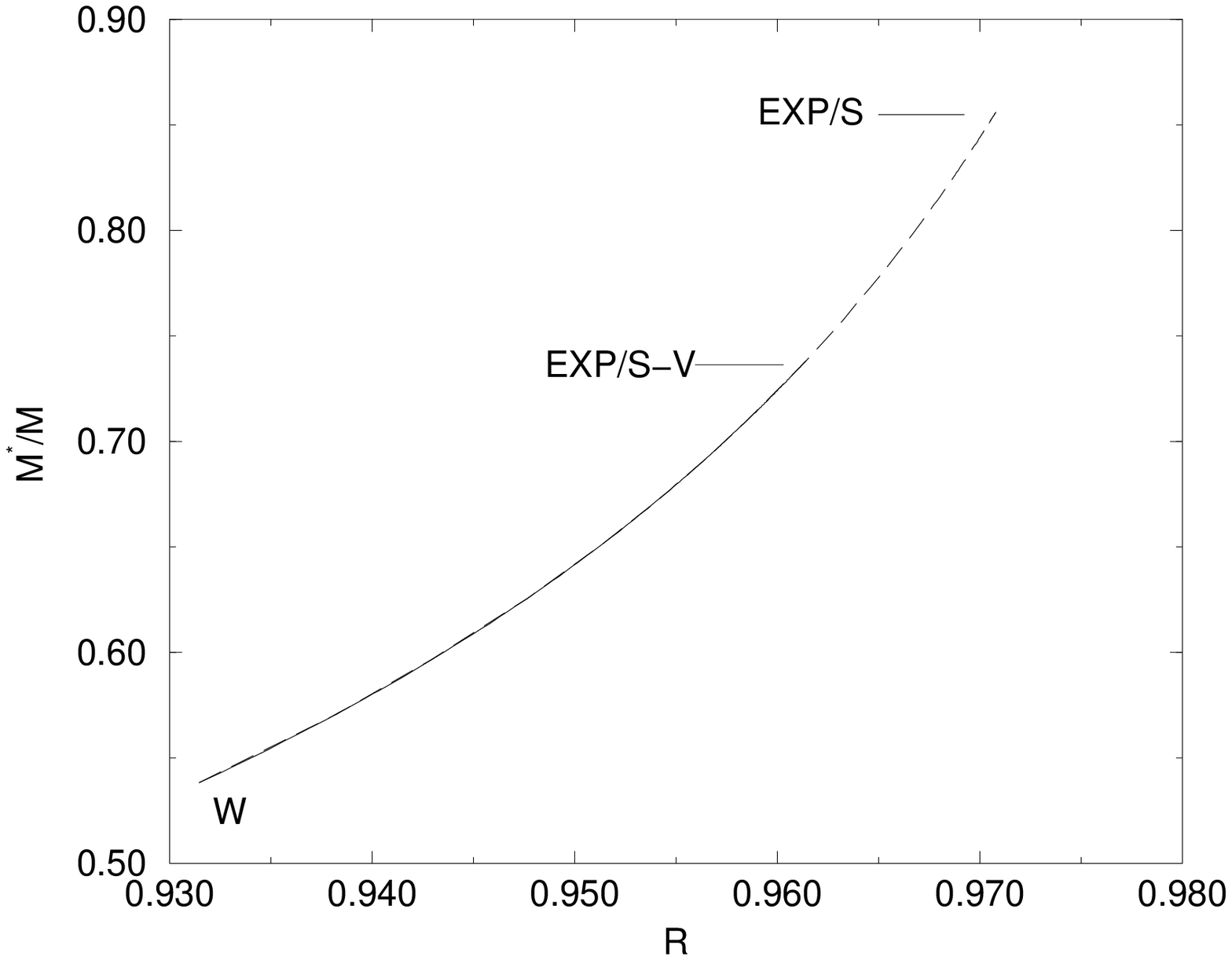}}
\end{figure}

\begin{figure}
\caption{Neutron star mass as a function of the central density in
Walecka (solid line), ZM (dashed line) and ZM3 (dotted line) models. \label{4zmotv2}}
\centerline{\epsfysize=8cm \epsfbox{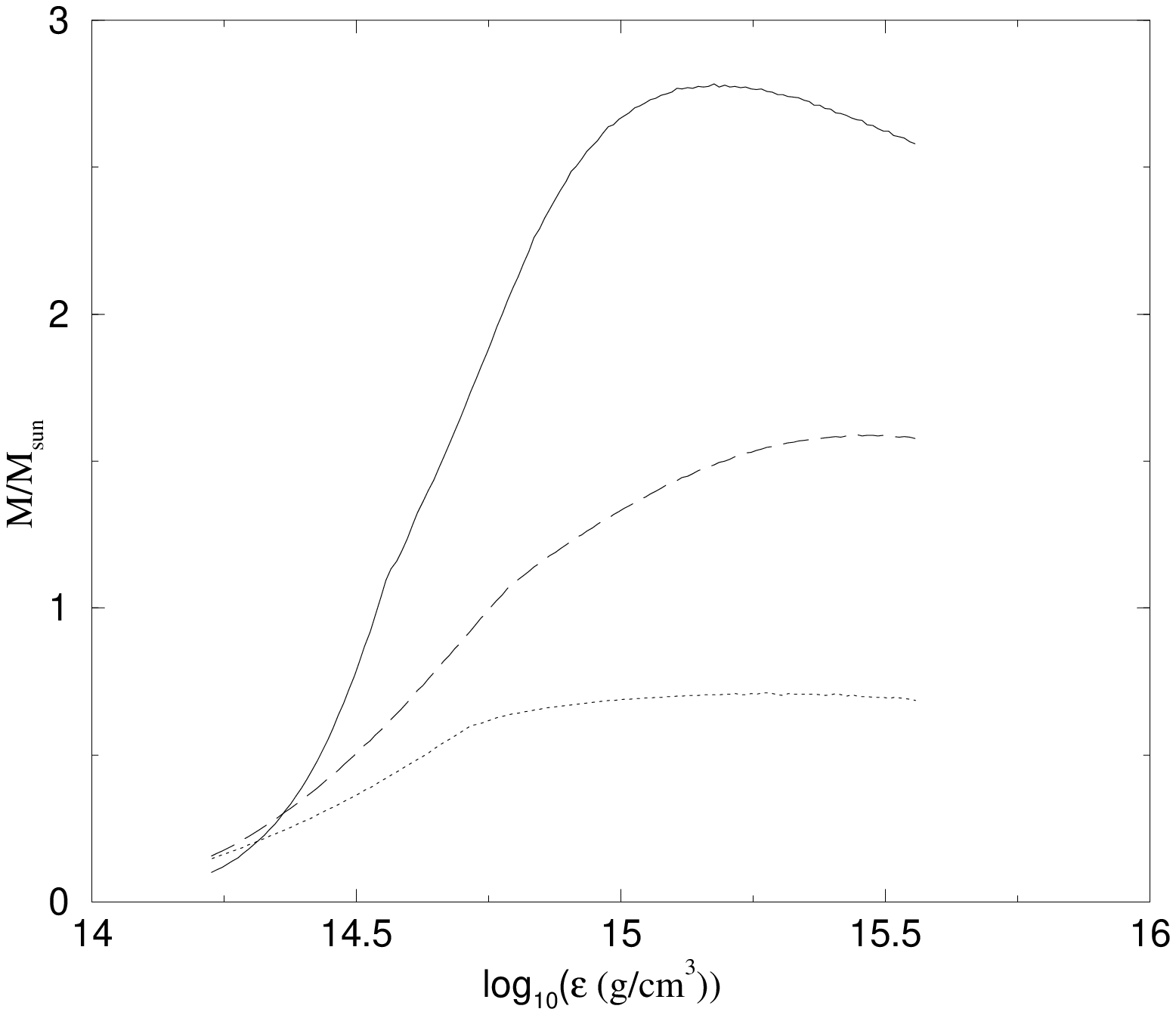}}
\end{figure}

\newpage
\begin{figure}
\caption{Chemical potentials and field intensities in the Walecka
({\em panel  a}), ZM ({\em panel  b}) and ZM3 ({\em panel c}) models .} \label{4zmpop2}
\begin{tabular} {c}
{\footnotesize{a)}} \\  \centerline{\epsfysize=6cm
\epsfbox{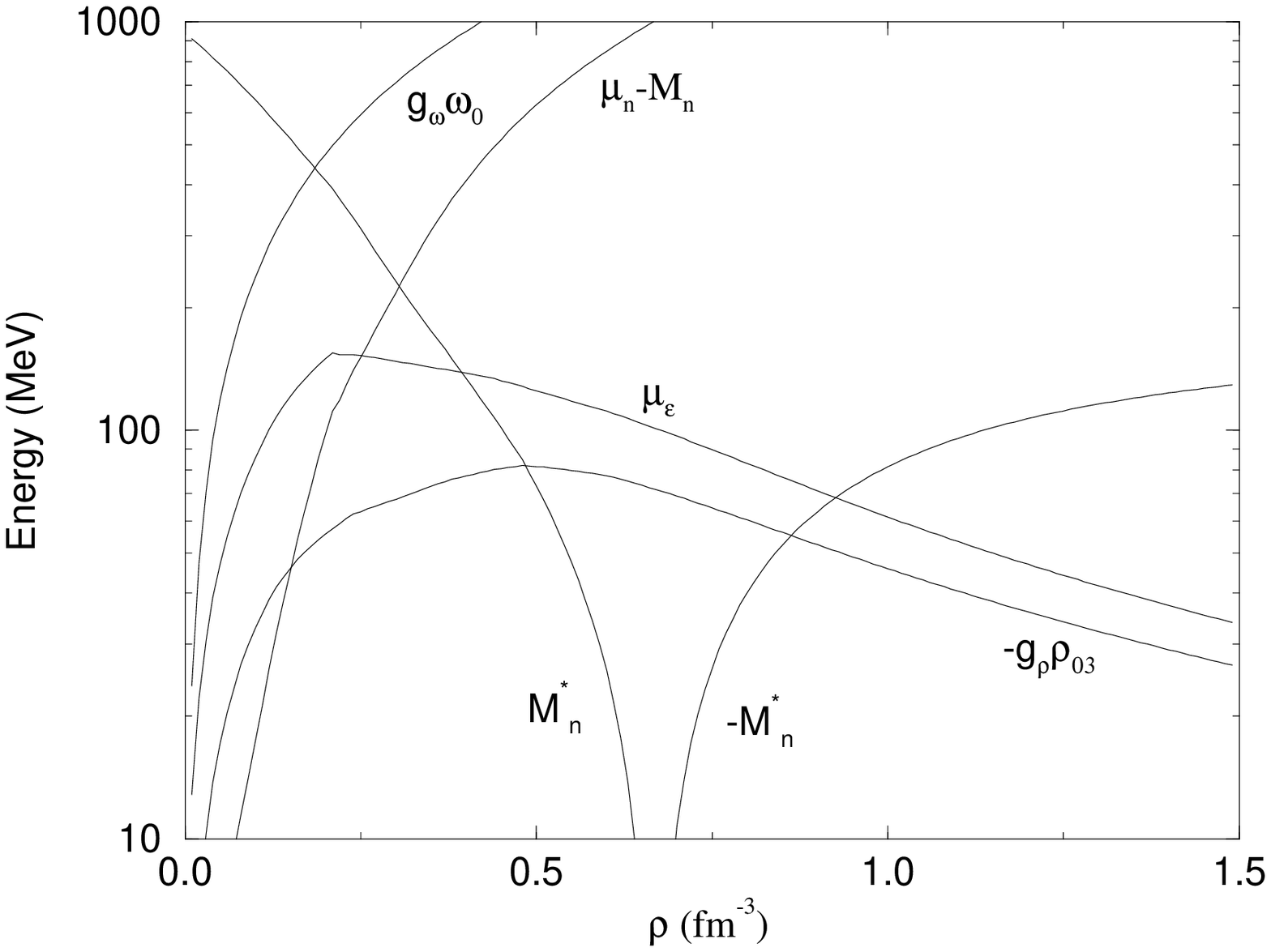}}
\\ {\footnotesize{b)}} \\ \centerline{\epsfysize=6cm \epsfbox{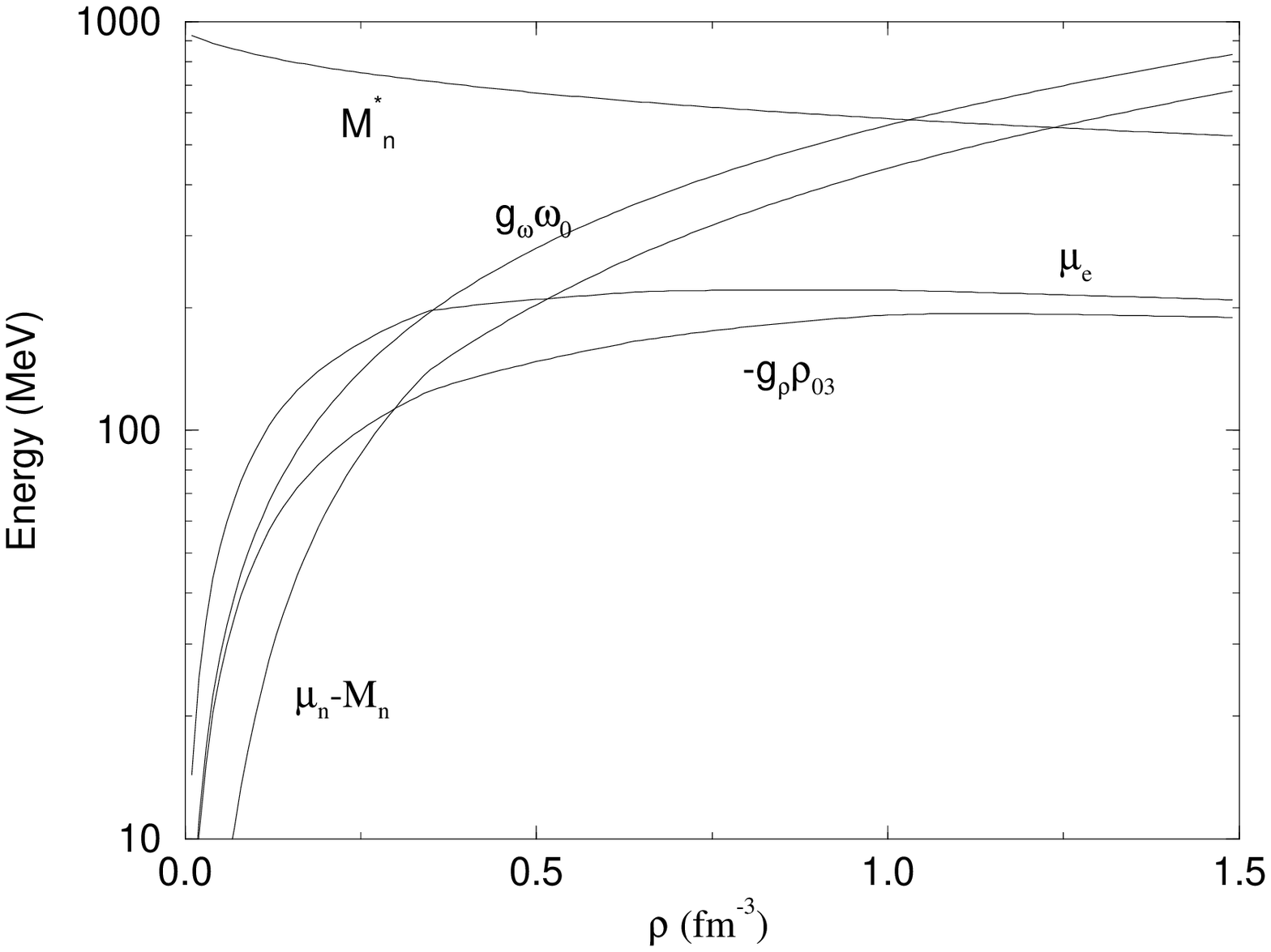}}\\
{\footnotesize{c)}} \\ \centerline{\epsfysize=6cm
\epsfbox{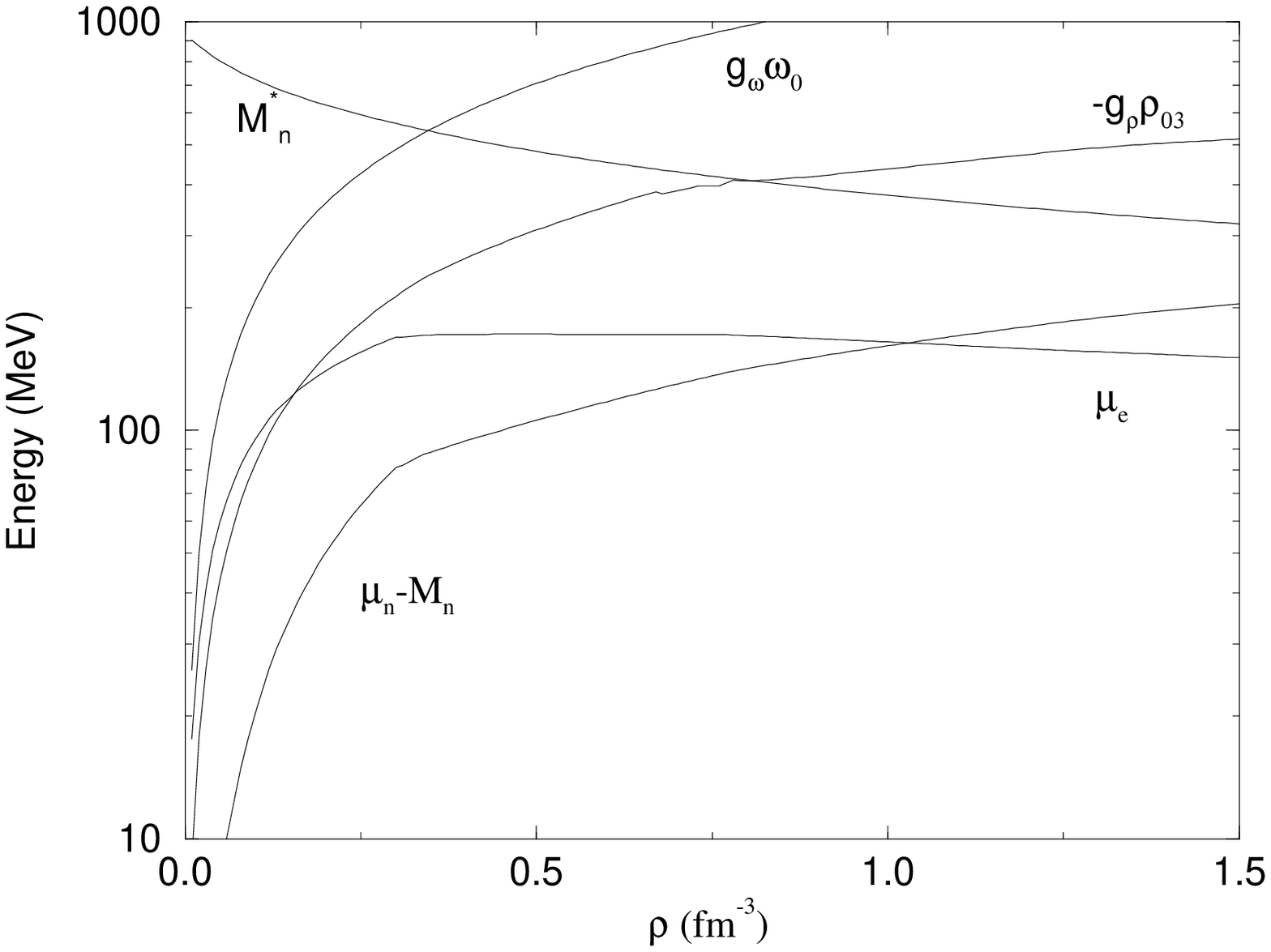}}
\end{tabular}
\end{figure}

\newpage

\begin{figure}
\caption{Baryon and lepton populations for the Walecka ({\em panel a})
and ZM ({\em panel b}) models.\label{fig4}}
\begin{tabular} {c}
{\footnotesize{a)}} \\  \centerline{\epsfysize=6cm
\epsfbox{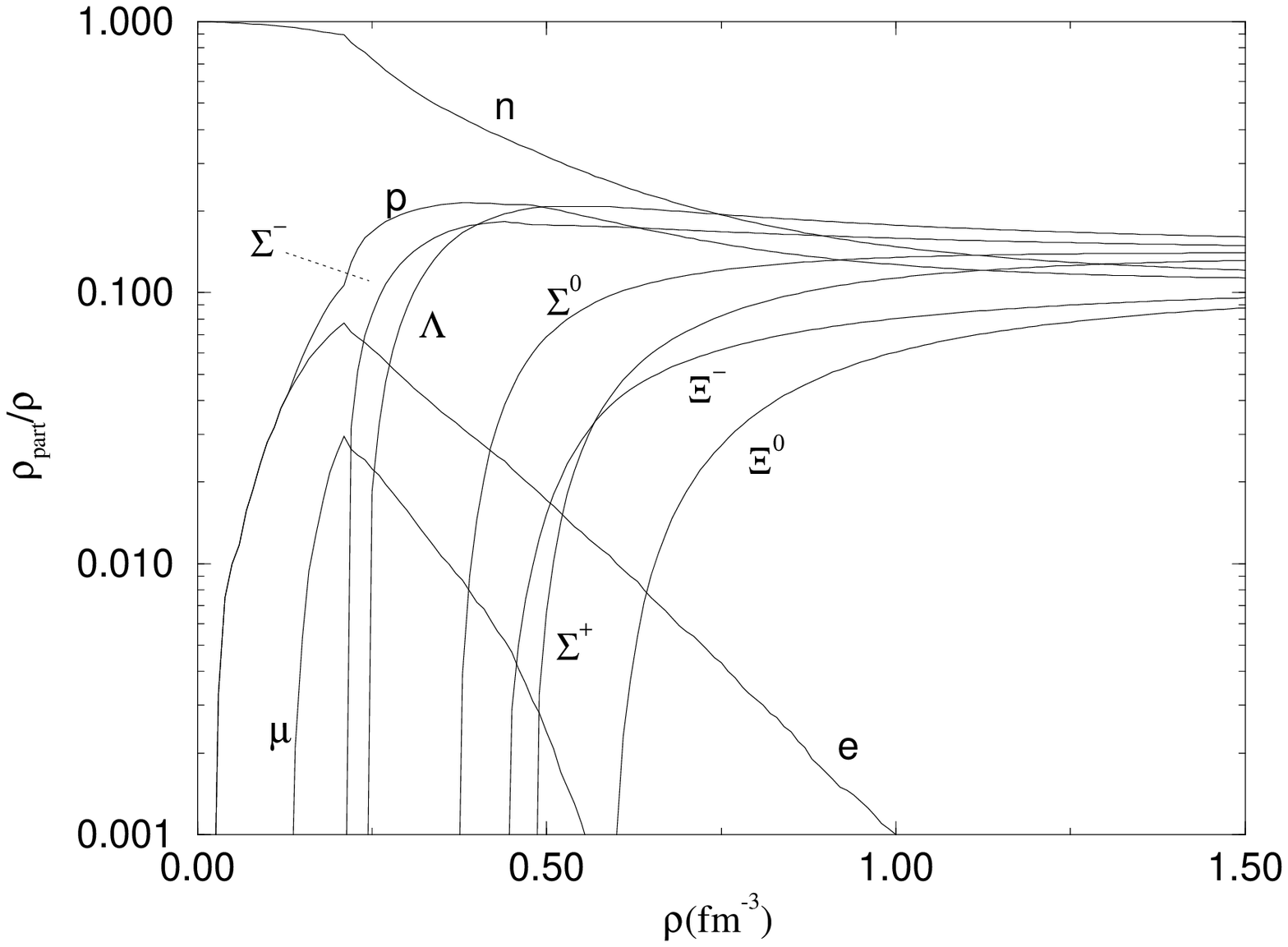}}
\\ {\footnotesize{b)}} \\ \centerline{\epsfysize=6cm \epsfbox{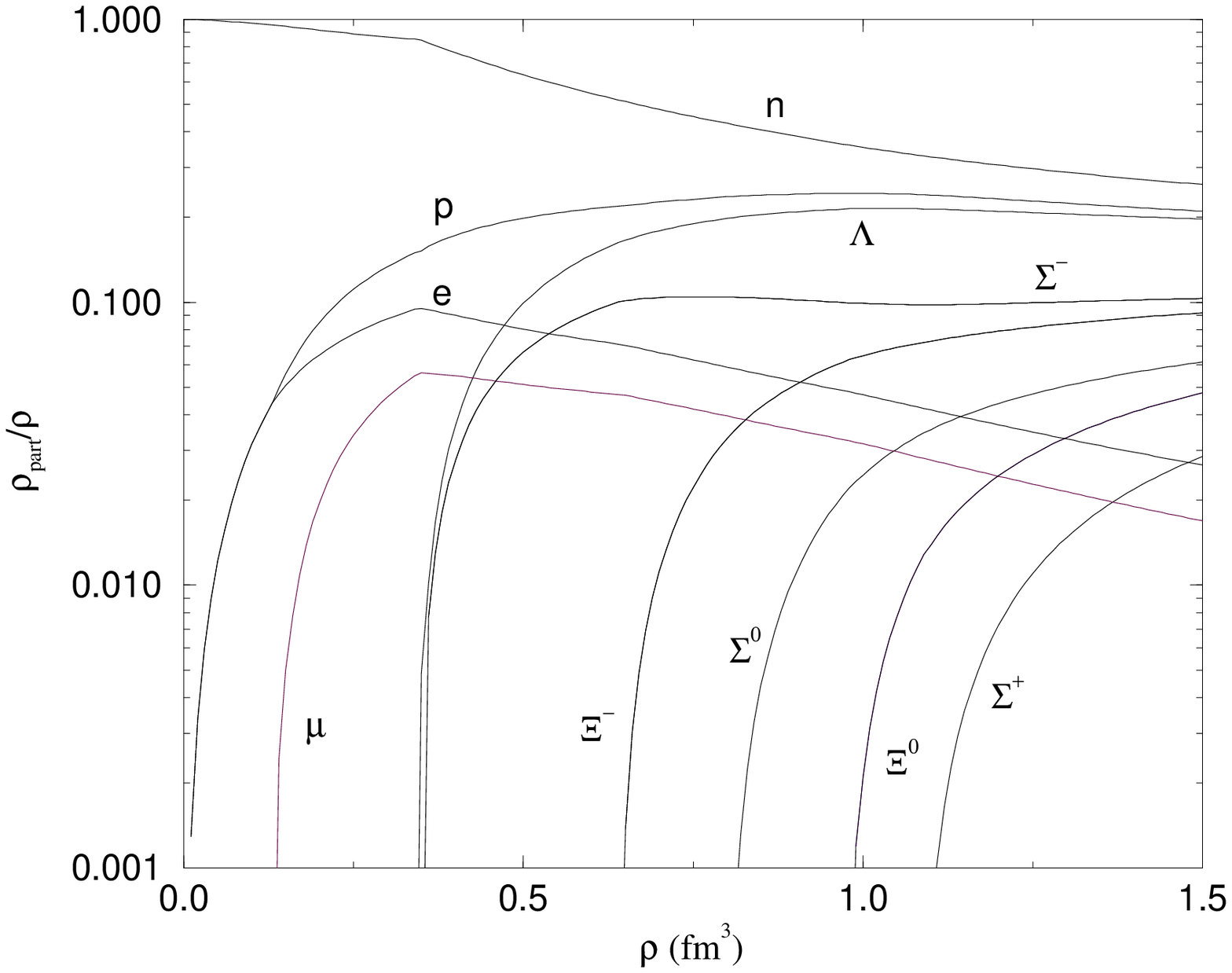}}
\end{tabular}
\end{figure}

\newpage

\begin{figure}
\caption{Radial
distribution of the different leptonic and baryonic species in the
ZM model. \label{4zmotv1}}
\centerline{\epsfysize=8cm \epsfbox{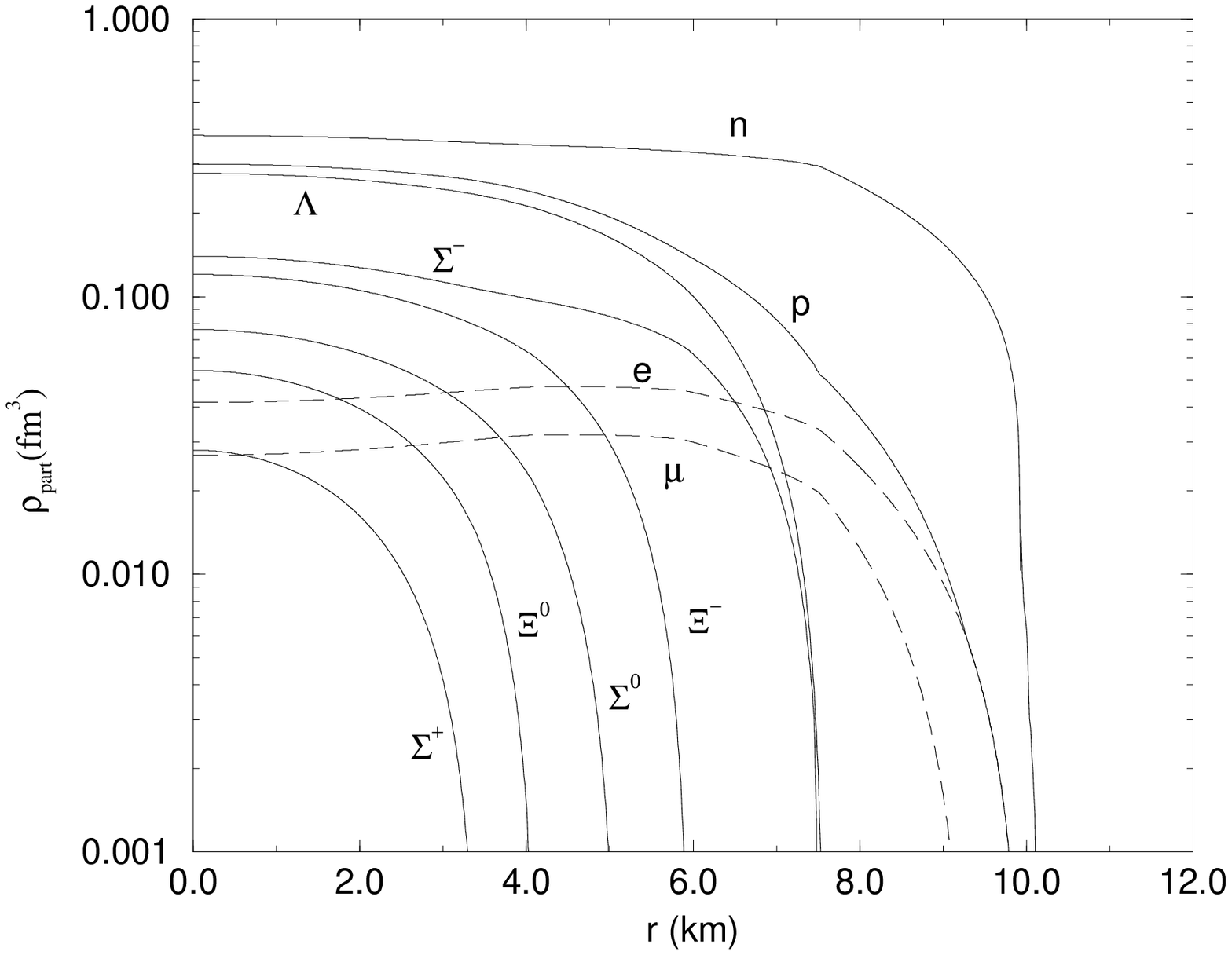}}
\end{figure}

\begin{figure}
\caption{Maximum mass of a neutron star
sequence (universal coupling) as a function of the $\lambda$
parameter, for cases S (full line) and S-V (long dashed line).}
\label{4mstlam}
\centerline{\epsfysize=8cm \epsfbox{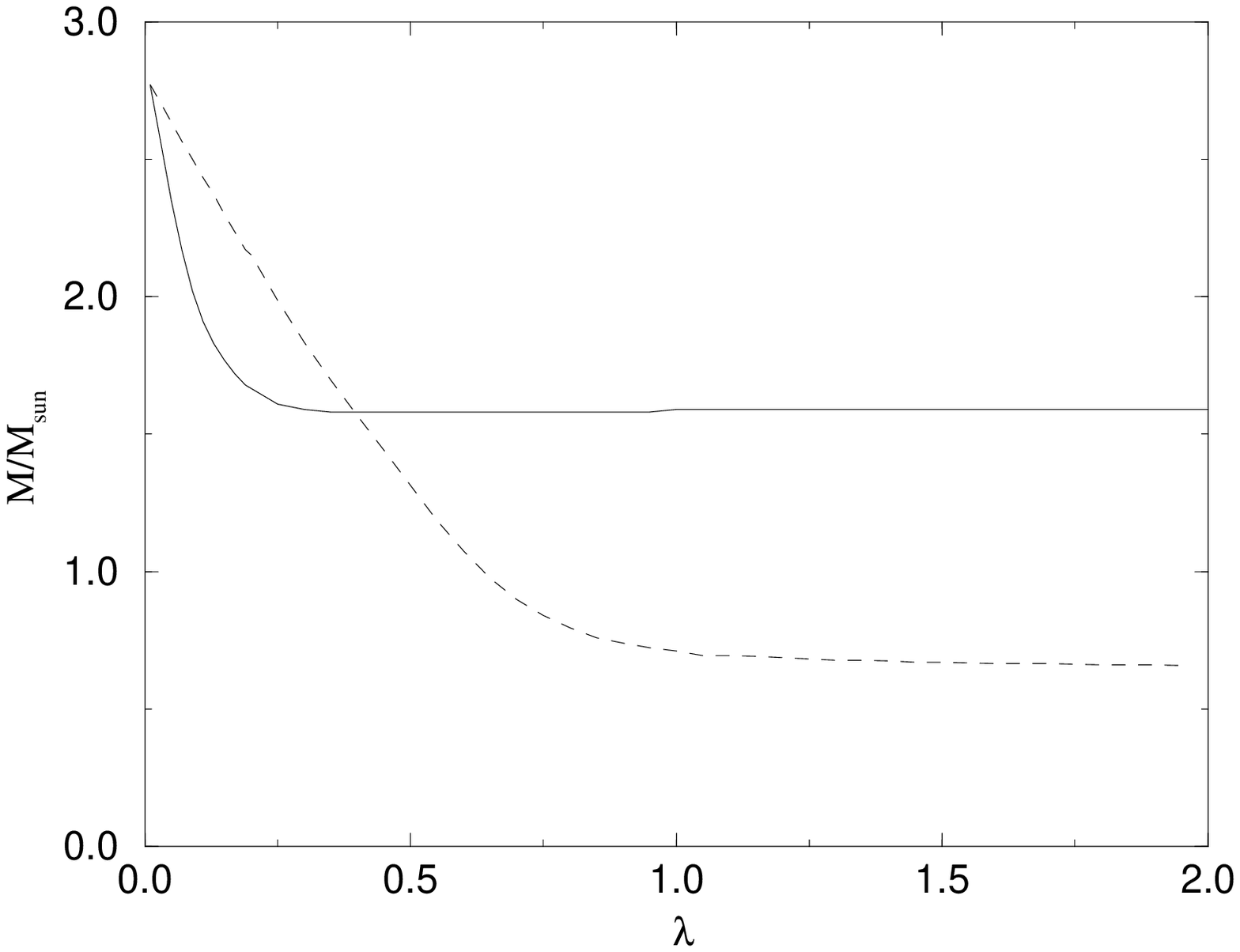}}
\end{figure}

\newpage

\begin{figure}
\caption{Dependence of the maximum neutron star mass of a
sequence with the compression modulus (left) and nucleon
effective mass at saturation (right). Solid line corresponds to
$S$ case and dashed line to $S-V$ case. \label{fig6}}
\centerline{\epsfysize=8cm \epsfbox{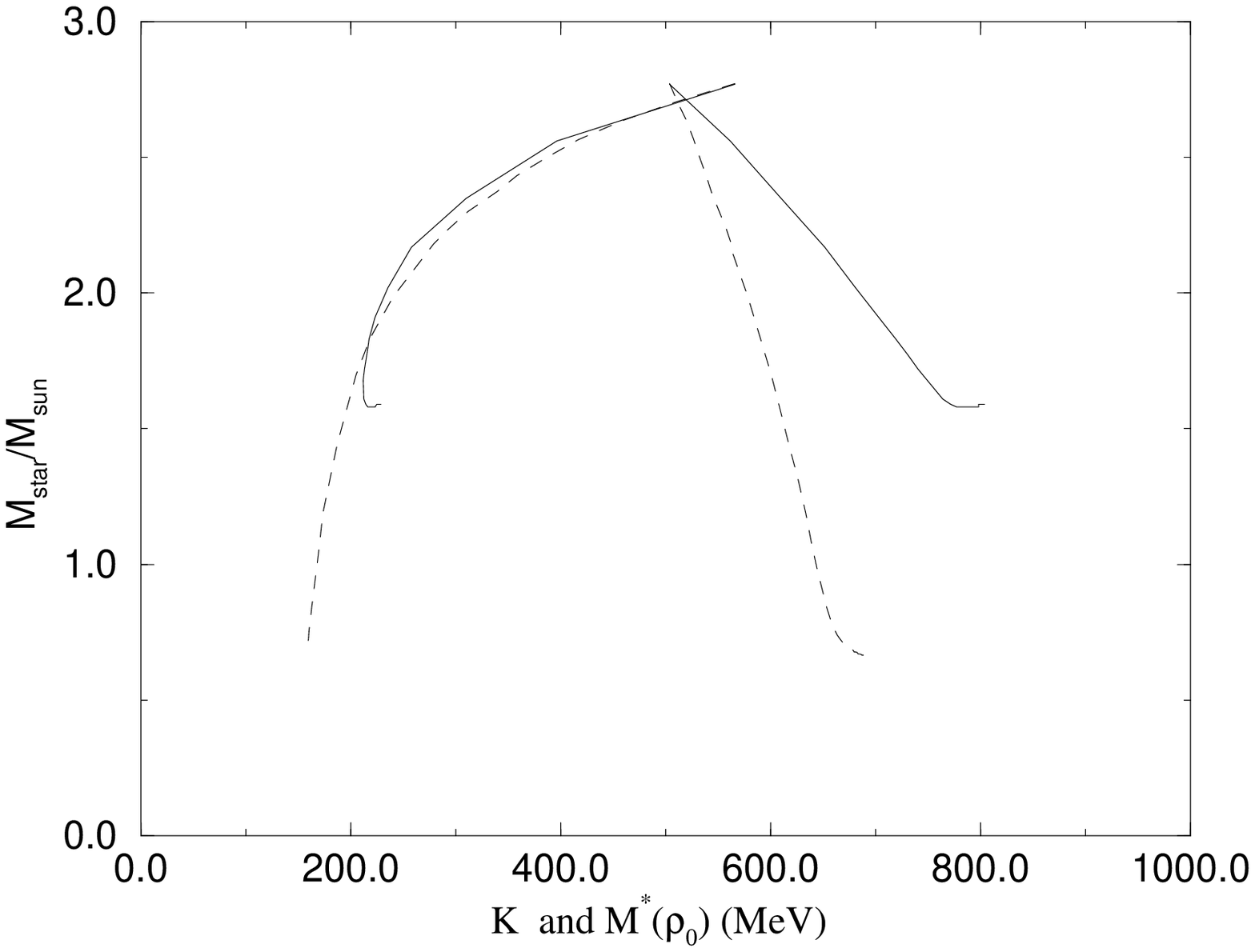}}
\end{figure}

\begin{figure}
\caption{Maximum neutron star mass as a function of the scalar
potential at the star center. Solid line corresponds to $S$ case
and dashed line to $S-V$ case. The long dashed line corresponds to
$S = 939$ MeV. \label{fig3}}
\centerline{\epsfysize=8cm \epsfbox{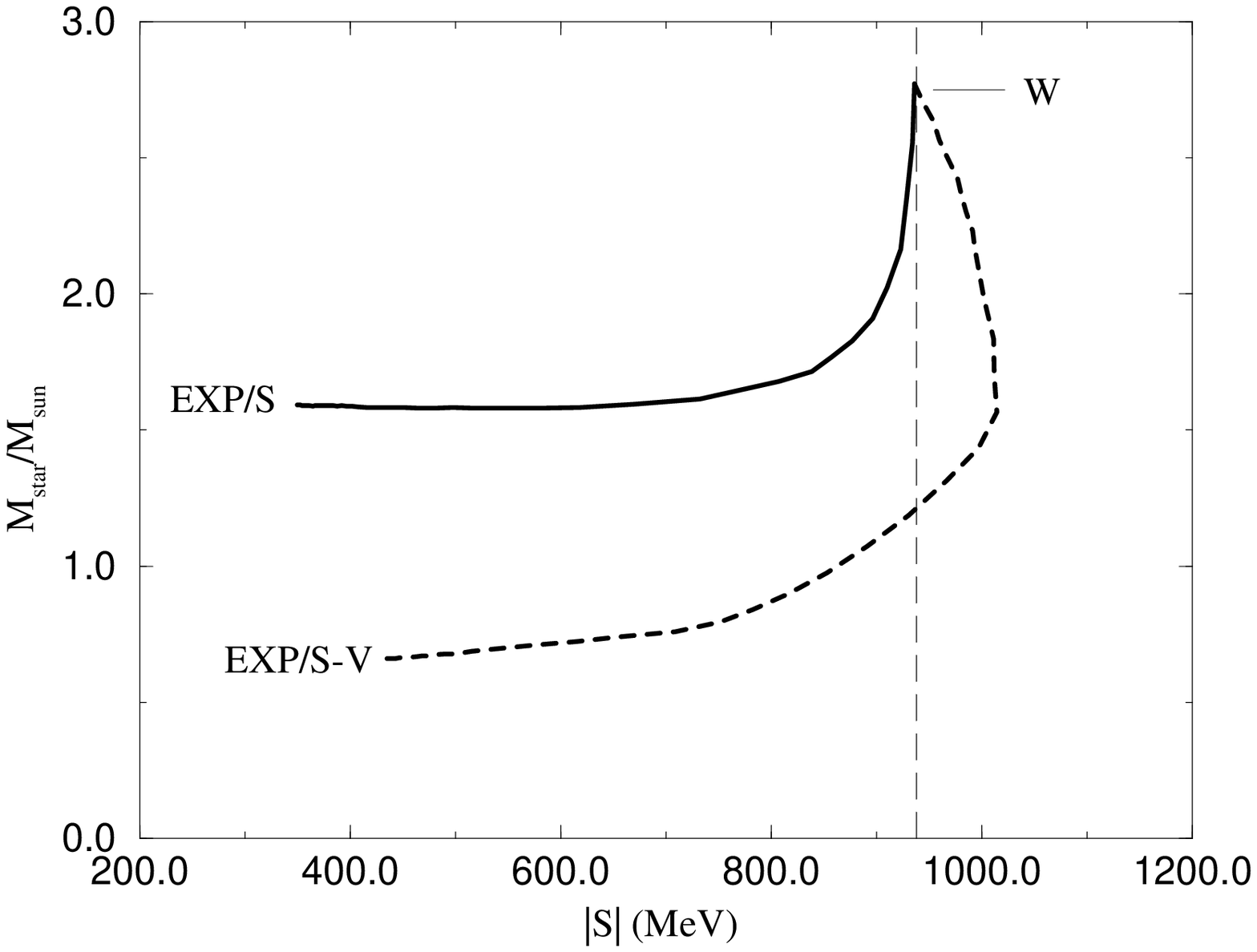}}
\end{figure}

\end{document}